\documentclass[]{aastex701}
\received{August 7, 2026}
\revised{September 9, 2026}
\accepted {September 21, 2026}
\submitjournal{AJ}

\usepackage{amsmath}
\usepackage{xcolor}
\usepackage{booktabs}
\usepackage{bm}
\usepackage[normalem]{ulem}
\defcitealias{jdox2016}{STScI, 2016}
\begin{document}

\title{Increasing Sensitivity to Trailed Solar System Objects in Archival JWST NIRCam Imaging with Group Differencing}

\author[]{Anthony Girmenia}
\affiliation{Western University, 1151 Richmond St, London, ON N6A 3K7, Canada}
\email{agirmen@uwo.ca}

\author[]{Stanimir Metchev}
\affiliation{Western University, 1151 Richmond St, London, ON N6A 3K7, Canada}
\affiliation{Western University, Institute for Earth and Space Exploration, London, Ontario, Canada}
\email{smetchev@uwo.ca}

\begin{abstract}
The JWST archive contains thousands of NIRCam exposures distributed across the sky. Small solar system bodies (SSSBs) are present in many of these fields, but are often trailed and more difficult to detect. The standard pipeline produces a single image by fitting a slope to a sequence of $N$ cumulative ``up-the-ramp" groups of the detector. Thus, each JWST image comprises a time series of cumulative reads. We demonstrate that transforming the ramp into a series of group difference images reduces the per-image trail length of moving sources, producing an $(N-1)$-group photometrically calibrated time series from an $N$-group integration. This enables higher-cadence trajectory-based searches using as few as a single JWST exposure. A generalized least-squares fit of the group difference images along the trajectory of a moving source recovers the source flux at higher SNR when compared to the ramp fit, with the improvement scaling with the reduction in trailing. We validate our procedure on 13 moving objects detected in archival JWST NIRCam images. To estimate our sensitivity increase, we perform an injection/recovery simulation on the JWST imagery. Trailed moving objects recovered via group differencing show higher SNRs, improving completeness to moving sources compared to standard calibrations. For typical integration lengths of $N=4$ or $6$ for archival JWST/NIRCam exposures in the ecliptic, group differencing yields sensitivity gains between $\sim 1.1$--$1.4$ mag, based on simulation results. This method significantly increases the sensitivity of archival JWST surveys to trailed SSSBs.
\end{abstract}


\section{Introduction}\label{sec:Intro}

Since beginning operations in 2022, the James Webb Space Telescope (JWST) has accumulated thousands of Near-Infrared Camera (NIRCam) images. Many of these observations are located at low ecliptic latitudes where the density of small solar system bodies (SSSBs) is highest. These fields consequently capture moving objects, such as main-belt asteroids, near-Earth objects, and trans-Neptunian objects. JWST/NIRCam's sensitivity, $0.6$--$5\,\mu\text{m}$ coverage, and simultaneous dual-filter capabilities cover a unique parameter space previously inaccessible without coordinated efforts across multiple observatories. The volume and diversity of these archival images offer a rich resource for characterizing both the physical properties and population statistics of various SSSBs. 

Several recent studies using JWST's Mid-Infrared Instrument (MIRI) demonstrated the potential of archival JWST images for both detecting and characterizing SSSBs. \cite{burdanovGPUbasedFrameworkDetecting2023} developed a GPU-based framework for identifying moving bodies in targeted exoplanet time-series imaging, and subsequently applied it to MIRI observations of the TRAPPIST-1 system, detecting $\sim140$ main-belt asteroids down to decameter sizes \citep{burdanovJWSTSightingDecametre2025}. \cite{mullerAsteroidsSeenJWSTMIRI2023} utilized MIRI images to derive radiometric diameters, distances, and orbital constraints for asteroids serendipitously present in the field of view. These past studies operate on integration-level calibrated products (one image per integration) and are restricted to mid-infrared wavelengths. A key obstacle, in contrast to rapid MIRI time-series observations, is that many NIRCam integrations are long enough for objects with moderate apparent motion to trail across multiple pixels. The standard JWST calibration pipeline collapses all temporal information within each integration into a single image via ramp fitting, making trailed sources difficult to detect. \cite{allenWaterIce} demonstrated that asteroids in NIRCam-pure-parallel fields \citep[GO 2211,][]{trillingPureParallelSurvey} can be identified by repurposing the jump-detection flags produced during standard ramp fitting. Pixels in which a moving object transits mid-integration are flagged as flux discontinuities, producing binary streak images. This approach operates within the existing calibration pipeline, but the jump flags provide detection positions rather than calibrated time series, and photometric extraction remains under development. Several accepted JWST programs have targeted asteroid detection in both parallel/archival NIRCam fields \citep[e.g., AR 3701, AR 8214][]{trilling_ar_2023, burdanov_ar_2025}, reflecting community interest in exploring this dataset. 

The previous approaches are limited by the representation of the temporal information within the ramp. As a moving object transits a pixel, it deposits signal over a time window shorter than the length of the integration. The source signal is spread over a larger spatial footprint, and when estimating the slope, the ramp-fit operation combines samples in which the source has already left the pixel. For an unbiased estimator, integrating the source's total flux over its trail can recover the true signal, but at the cost of increased background and read noise from the larger effective aperture. Alternatively, the jump-detection approach identifies moving sources by detecting inter-group flux discontinuities, but does not preserve the underlying flux measurements as a calibrated time series. Thus, while these approaches utilize aspects of the temporal information encoded in the ramp, the full source, noise, and temporal information available within an integration has not yet been leveraged for moving-object photometry and detection.

In this work, we apply the group-difference representation of up-the-ramp-data \citep[][]{Kubik_2015} to JWST moving-object detection and photometry. By transforming each NIRCam integration into a sequence of short-exposure group-difference images, we recover the intra-integration temporal and photometric information that is otherwise compressed by standard ramp fitting. Each difference image has a shorter exposure time that reduces the trailing losses for objects that move $\geq1$ PSF FWHM within individual integrations. The scientific potential of intra-exposure ramp data for moving-object detection has recently been recognized in the context of the Roman Space Telescope \citep{RampsEnablingSolar2026}, but it has not yet been implemented in practice, to the best of our knowledge. The method transforms single, long-integration NIRCam images into a uniform time series, recovering the temporal/photometric information that ramp fitting destroys. The resulting frames are fully calibrated flux-rate images suitable for standard photometric and astrometric measurements. Because moving sources appear in different positions each frame, detections across the image sequence can be linked into intra-integration tracklets, providing both motion confirmation and false-positive rejection within single integrations. The intra-exposure sequence provides additional frames compared to the default calibrated sequence. This enables more robust applications of trajectory-based moving object searches, targeting populations with greater on-sky angular velocities and allowing for more independent measurements to be stacked.

The remainder of this work is organized as follows. Section \ref{sec:methods} describes the group difference calibration procedure. Section \ref{sec:observations} presents a validation of the technique on 13 moving objects detected in archival JWST/NIRCam imagery and reports the results of an injection/recovery simulation used to estimate the sensitivity increase gained from group differencing over the standard ramp-fit. Section \ref{sec:discussion} discusses the archival opportunity, applicability to other instruments, and limitations. Section \ref{sec:conclusion} summarizes our work. 

\section{Difference Image Representation of a Ramp} \label{sec:methods}

\subsection{NIRCam's ramp structure and its effect on moving sources}\label{sec:nircam-structure}

All NIRCam readout patterns consist of a series of non-destructive reads of each pixel, each referred to as a ``frame". Depending on the readout pattern, frames are averaged into ``groups" of $m$ frames, with the possibility of several frames being skipped between groups. A set of groups comprises an ``integration" or ``ramp". This follows the established JWST NIRCam terminology laid out in the JWST documentation \citepalias{jdox2016}. The majority of available readout patterns do not preserve the individual frames; therefore, the group sequence is the finest temporal sampling generally available. {For convenience, the quantities used throughout this section are summarized in Table \ref{tab:notation}.}

\begin{table}
    \centering
    \begin{tabular}{c|c}\toprule
        Symbol & Definition \\\midrule
        $T_{\rm int}$ & Integration effective exposure time \\ 
        $G_k$ & $k$th group measurement \\ 
        $N$ & Number of groups per integration \\ 
         $L_\text{int} $ & Trail length of moving source in ramp-fit image  \\
         $L_G$ & Trail length of moving source in group difference image \\
         $D_{kk}$ & Variance of the $k$th group difference in a ramp \\
         $D_{kl}$ & Covariance between group differences $k$ and $l$ in a ramp\\
         $f$ & Source count rate \\
         $b$ & Background count rate per pixel \\
         $r$ & Read noise per pixel \\
         $n_{\rm pix}$ & Number of pixels contained within the photometric aperture\\\bottomrule
    \end{tabular}
    \caption{{Notation used in Section \ref{sec:methods} and their definitions.}}
    \label{tab:notation}
\end{table}

\begin{figure}
    \centering
    \includegraphics[width=\textwidth]{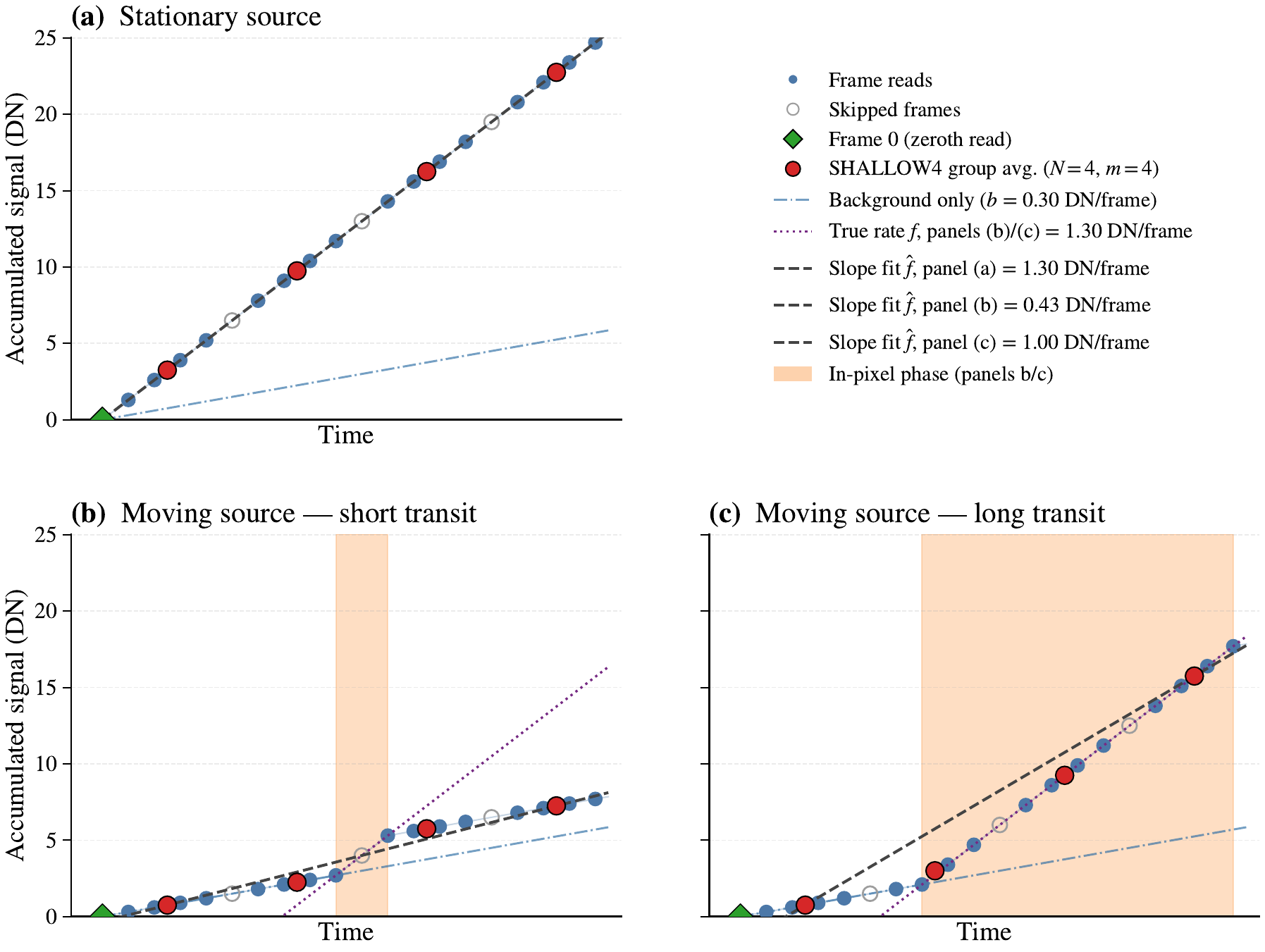}
    \caption{Up-the-ramp slope fitting applied to a stationary vs. a moving source, illustrated for the NIRCam SHALLOW4 readout pattern ($N=4$ groups, each averaging $m=4$ frames with one frame skipped). Read noise is neglected in this example. {(a)} For a stationary source, the accumulated charge rises linearly with time, and the slope fit through the group averages (red points) recovers the true count rate exactly. {(b)} The same fit applied to an asteroid that transits the pixel within a single inter-group interval (orange band). {(c) A more general case in which the source transit overlaps multiple groups {(in this case three)} and their frame-averaging windows, distributing the source signal across multiple adjacent group differences.} $f$ is the true count rate of the asteroid, while the estimator $\hat{f}$ underpredicts the true rate due to the source only being present in the pixel for a limited time. The additional noise contribution by the background from epochs where the source was not present in the pixel degrades the SNR of moving objects in standard full-ramp calibrated products. In the background-noise regime, the slope-fit weights collapse to the difference between the last and first groups in the ramp, photometrically equivalent to a destructive CCD read of the same length.} 
    \label{fig:nircam-ramp-schematic}
\end{figure}

\begin{figure}
    \centering
    \includegraphics[width=\textwidth]{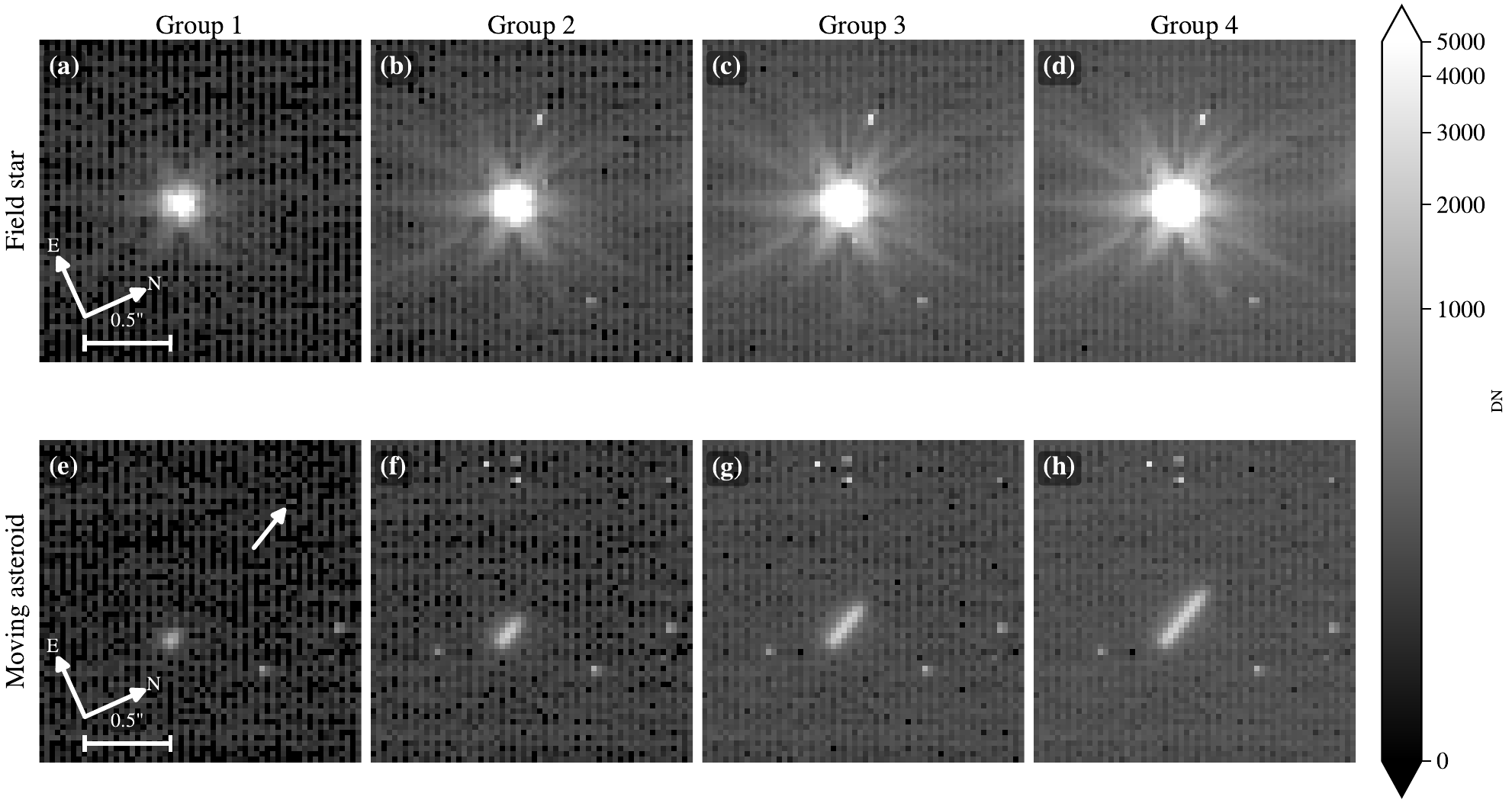}
    \caption{Group reads for a stationary field star (top) vs. for a serendipitously detected numbered asteroid 417988 (bottom). Images are taken from JWST PID 5594. These images contain a ramp of $N=4$ groups. The pixel scale of each cutout is $0.031\arcsec$/pixel. The default ramp-fit assumes a stationary source, whereas a moving source will cause the slope in each pixel it transits to be underestimated when compared to the asteroid's true flux rate.}
    \label{fig:4-panel-field-star-vs-asteroid}
\end{figure}

The standard JWST calibration pipeline produces a single image per integration by fitting an optimally weighted least-squares slope to the accumulated signal in each pixel as a function of time. This process assumes constant illumination, modelling every pixel as a static source with a fixed count rate. A moving source violates this model by illuminating a given pixel only for the fraction of the integration during which the source's PSF overlaps it. The slope fit is then applied to a transient bump, rather than to a ramp with a constant slope. The fit of each pixel will respond by returning a suppressed count rate, with the remainder of the object's flux being redistributed along a trail of length $L_{\rm int}=(N-1)L_G$ rather than within a compact PSF. While trailing is common to any imager with a finite exposure, for an up-the-ramp detector, the trail profile is set in detail by the slope estimator's weights. Figure \ref{fig:nircam-ramp-schematic} compares the ramp for a stationary source against an asteroid moving across the NIRCam detector during a multi-group integration. Figure \ref{fig:4-panel-field-star-vs-asteroid} plots the individual group reads for a stationary field star along with a moving asteroid detected in JWST imagery.

A caveat of the standard JWST pipeline is the \texttt{jump-detection} step, which is intended to remove cosmic ray artifacts from the imagery by detecting and subtracting flux discontinuities in the ramp. This step is known to mask moving objects and is trivially turned off by changing the default parameters of the pipeline.

In the photon-noise-limited regime (including both background-noise and source-noise), the optimal up-the-ramp slope estimator used by JWST/NIRCam reduces to the two-point difference between the first and last groups of the ramp \citep[][]{bushouse_2026_20058613}. The recovered count rate in each pixel is then
\begin{equation}
    \hat{f} = \frac{G_{N-1}-G_0}{T_\text{int}}.
\end{equation}
where $\hat{f}$ (counts/second) is the estimated count rate, $G_i$ (counts) is the $i$th group read, and $T_\text{int}$ (seconds) is the total exposure time of the integration. Even in the case of a transient source, $\hat{f}$ implies that the charge that was accumulated over the integration is exactly the total charge deposited in the pixel. In other words, it is photometrically equivalent to a finite exposure with the same duration $T_{\rm int}$.

\subsection{Group differencing as intra-integration digital tracking}\label{sec:digitaltracking}
For a source with count rate $f$ (counts/second) measured against a per-pixel background $b$ (counts/second/pixel) {with read noise $r$ (counts/pixel)}, the SNR is given by 
\begin{equation}\label{eq:snr}
    {\mathrm{SNR} = \frac{ft}{\sqrt{ft+n_{\rm pix}(r^2+bt)}}}
\end{equation}
where $n_{\rm pix}$ is the effective number of pixels over which the detection accumulates background noise. A trailing source will spread its flux $f$ across $L_{\rm int}=(N-1)L_G$ pixels over the course of an integration, and will thus have $n_{\rm pix}\propto t$. Any detection aperture applied to the rate image must accumulate the background along the full footprint $L_{\rm int}$, limiting the effective exposure time over which $\mathrm{SNR}\propto\sqrt{t}$ to the time in which the source remains compact. Digital tracking techniques \citep[e.g.,][]{heinzeDigitalTrackingObservations2015} provide a framework for reducing this SNR loss by instead taking shorter exposures to reduce the size of the velocity-space over which sources trail, and shift-and-stacking along a moving-objects trajectory. This procedure accumulates the same signal $f$, but over a smaller {detector footprint}. 

\begin{figure}
    \centering
    \includegraphics[width=\textwidth]{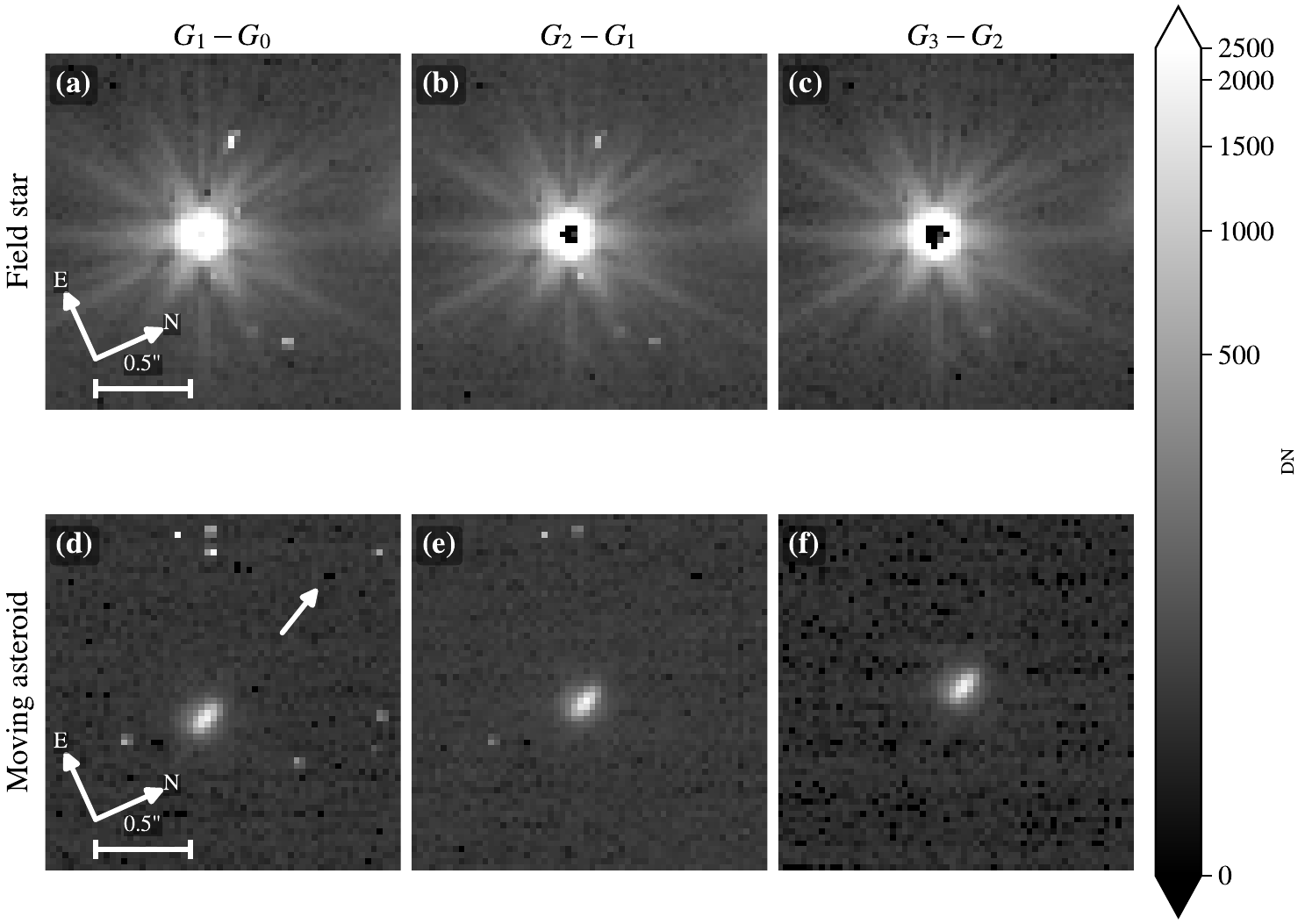}
    \caption{The sequence of $N-1$ group differences obtained from successive ramp fits to adjacent pairs of group reads, obtained from the same ramp shown in Figure \ref{fig:4-panel-field-star-vs-asteroid}. The trail length of the asteroid in each group difference is shorter than its length in the full ramp fit, allowing one to integrate over fewer background pixels using shift-and-stack procedures. Group differencing has the additional benefit of increased temporal sampling of the ramp, allowing for the motion of a candidate asteroid to be confirmed within single JWST exposures. Time series of higher cadences than what is possible with the full ramp-fits can also be constructed for both stationary and moving targets in the images. The dark black pixels in the center of the field star's PSF are due to early saturation, such that successive group differences oversubtract the stellar core.}
    \label{fig:ramp-slice-example}
\end{figure}

Working in the group-difference representation of the ramp enables this technique to be applied post-hoc to archival JWST imagery, including single integrations, by extracting the higher-cadence intra-exposure group reads from the ramp, and recalibrating them into intra-ramp count-rate images. Each of the $N-1$ differences captures the moving source over a single inter-group interval, during which it crosses only $L_G$ pixels. Digital tracking techniques can then be realized within single JWST integrations. The magnitude of this gain depends on the original size of the integration/ramp and on the duration of a transiting source (or a transient event) in a pixel, as longer ramps will have caused a larger range of velocities to trail such that additional exposure no longer improves their detectability. Figure \ref{fig:ramp-slice-example} plots the $N-1$ group differences obtained from applying the procedure to the same ramp shown in Figure \ref{fig:4-panel-field-star-vs-asteroid}.

\subsection{{Effect of transit timing on group-difference photometry}}

{Figure \ref{fig:nircam-ramp-schematic} illustrates how the timing of a source transit relative to the group measurements determines both the calculated ramp-fit slope and the source's representation in the group differences. In the simplest case, the source transits a pixel entirely between two successive groups, such that the corresponding difference measures the charge accumulated between the group measurements. More generally, however, the transit may overlap one or more group-averaging windows. Because each group represents the average of $m$ individual frames, a group difference does not necessarily isolate the charge deposited strictly between the nominal group measurement times. Instead, the measured signal reflects the source's presence within the constituent frames of both groups. Consequently, when a source is present during a group-averaging window, its signal can be distributed between the two neighbouring differences that share that group.  This changes how the signal appears in individual group differences but does not change the total recovered source signal when the affected differences are combined.}

{The same averaging also affects the spatial footprint of a moving source in an individual group difference. The region containing the source signal is not limited to the displacement between the nominal times of two successive groups, but can span the trajectory from the source position in the first frame of the first group to that in the final frame of the final group. An aperture encompassing this full trajectory therefore captures all pixels that can contain source signal. For the uniform NIRCam readout patterns considered here, both the frame-averaging pattern and inter-group spacing remain constant throughout an integration. Assuming constant angular velocity, the required aperture has the same dimensions in every group-difference image and can simply be translated along the source trajectory. For instruments with non-uniform sampling, such as the Roman space telescope, the required aperture extent would instead vary between differences.}

\subsection{Noise properties of group differences and their co-addition}\label{sec:slicenoise}

Straight-line fits to differenced groups, including their exact noise properties under $m$-frame averaging, were analyzed by \cite{Kubik_2015}, whose covariance formulae we adopt below. The group-difference set is a reparameterization of the ramp, and contains all the same information and noise properties.

Successive group differences are not statistically independent. Adjacent group differences share one group, whose read noise and Poisson fluctuations enter both. For a NIRCam readout pattern with $m$ averaged frames per group and $n_{\rm skip}$ frames skipped between groups, the per-pixel covariance of the difference for a constant rate $f$ follows directly from the group-difference covariance of \cite{Kubik_2015} Eqs. 6--7:

\begin{equation}\label{eq:diagonal-cov}
    D_{kk} = \frac{2\sigma_R^2}{m} + d + \frac{(m-1)(2m-1)}{3m}f,
\end{equation}
\begin{equation}\label{eq:off-diagonal-cov}
    D_{kl} = \frac{(m^2-1)}{6m}f-\frac{\sigma_R^2}{m}
\end{equation}
with all covariances beyond adjacent differences, i.e., for $|k-l|>1$, being zero. Here, $\sigma_R$ is the per-frame read noise, such that $\sigma_R^2/m$ is the effective per-group read-noise variance after the averaging of frames. $f$ is the average inter-frame flux, and $d$ is the average inter group flux. We note that the first term of Equation \ref{eq:diagonal-cov} of \cite{Kubik_2015} appears to contain a typographical error, with the read-noise term written as $2\sigma_R/m$ rather than as the read-noise variance $2\sigma_R^2/m$. We adopt the latter form here, which is consistent with the definition of $\sigma_R$ as the per-frame read-noise standard deviation. 

The set of group differences forms a complete, invertible basis for the pedestal-free component of the ramp, preserving all temporal and photometric information relevant to linear slope estimation. For stationary sources, a generalized least-squares combination of the group differences using their covariance matrix recovers the optimal ramp-fit slope at a similar SNR \citep[][]{Kubik_2015}. For moving sources, the same formalism can be extended by measuring the aperture-integrated source flux within each group difference, with the covariance matrix providing the appropriate weighting for combining the correlated measurements across group differences. We further note that a useful simplification exists for rapidly moving sources whose displacement exceeds $\sim 1$ PSF FWHM between successive group-difference images. In this regime, the source occupies a distinct set of detector pixels in each difference image, so the extracted flux measurements no longer share detector pixels and can be treated as independent.

\section{{Validation with Archival NIRCam Imaging: PID 5594}}\label{sec:observations}

We aim to validate our calibration procedure on serendipitously detected moving objects in archival JWST/NIRCam imagery. We searched the Barbara A. Mikulski Archive for Space Telescopes (MAST) archive to find sample datasets with $N\geq3$ groups per integration focused near the ecliptic plane, so that the standard ramp-fitting and group differencing approach can be compared. We tested our approach using archival images from the JWST cluster SLICE survey (PID 5594). This program was an extragalactic NIRCam survey observing with F150W2+F322W2 from cycle 3. PID 5594 was chosen for its homogeneity: each pointing yielded 27 group-difference images with a fixed readout pattern and filters. All the JWST data from PID 5594 can be found in MAST: \dataset[10.17909/w611-5v71]{http://dx.doi.org/10.17909/w611-5v71}.

\subsection{Calibrating JWST/NIRCam Group Difference Frames}

We selected the 14 pointings of PID 5594 with {ecliptic latitude} $|\beta|\lesssim10^\circ$ out of the 124 available. A summary of our pilot dataset is shown in Table \ref{tab:Go-5594-summary}. We retrieved 630 \texttt{*uncal.fits} files from MAST (9 exposures $\times$ 14 pointings $\times$ 5 detector modules per pointing) and ran the STScI \texttt{Detector1Pipeline} on each file up to the \texttt{RampFitStep}. Ramp fitting was then looped over successive pairs of adjacent groups using the \texttt{--first\_group} and \texttt{--last\_group} flags. As a straight line fit between two points is unique, this is equivalent to creating group difference images and normalizing by the inter-group exposure time. The \texttt{jump} step was disabled, as cosmic ray hits are confined to single images (additionally, with only two groups in the ramp, there is no baseline). We saved the results as custom \texttt{*rategroups.fits} files analogous to the standard \texttt{*rateints.fits} files produced by the standard pipeline.   Each \texttt{*rategroups.fits} file was then processed through the STScI \texttt{Image2Pipeline}, producing fully calibrated \texttt{*calgroups.fits} files suitable for source detection and astrometry. A summary of the program and its calibration is displayed in Table \ref{tab:Go-5594-summary}. {The group-difference calibration procedure described here was implemented using the \texttt{JWST-Ramp-Slicer} Python code \citep{rampSlicerv1.1.1}. }

\begin{table}
    \centering
    \begin{tabular}{c|c}\toprule
         Program ID & 5594\\
         PI & Guillaume Mahler \\
         Filters & F150W2+F322W2 \\
         Read mode & SHALLOW4 \\
         Groups per integration & 4 \\
         Integrations per exposure & 1 \\
         Number of exposures per pointing & 9 \\
         Number of group difference frames per pointing & 27 \\
         Stare time per pointing & 1836 s \\
         Number of pointings with $|\beta|\lesssim 10^\circ$ & 14 \\
         Mean $\beta$& $4.9^\circ$     \\
         Total area covered ($|\beta|\lesssim 10^\circ$) & 72.6 square arcminutes \\\bottomrule
    \end{tabular}
    \caption{Summary of JWST program PID 5594. }
    \label{tab:Go-5594-summary}
\end{table}

\subsection{Measuring candidate movers}

We identified candidate moving objects in our observations using the Tycho Tracker software package on our group-difference images \citep[][]{parrottTychoTrackerNew}. Tycho Tracker was previously applied to archival JWST searches by \cite{burdanovGPUbasedFrameworkDetecting2023}. Although Tycho Tracker does not perform covariance-aware co-addition of the group-difference frames, for trailed sources, few to no detector pixels are shared by the source's footprint between group difference frames, and thus they can be treated as approximately independent. Developing a fully covariance-aware blind shift-and-stack pipeline is beyond the scope of this work. Candidate sources were vetted using Tycho Tracker's built-in classifier, as well as by eye to verify $\geq 3$ observations of coherent motion. Post-detection, photometry was performed using either circular apertures for untrailed sources, or rectangular apertures for trailed sources, aligned with the source's trail length and direction. {The apertures used for source photometry ranged in size from $4\times4$ pixel circular apertures for effectively untrailed sources to $4\times22$ pixel rectangular apertures for strongly trailed sources.} The aperture sum of a source in a group difference frame is given by Equation \ref{eq:aperture-sum}
\begin{equation}\label{eq:aperture-sum}
    A_i = \sum_p w_{i,p}I_{i,p},
\end{equation}
where $A_i$ is the aperture sum of the source in the $i$th difference image, $w_{i,p}$ is a binary mask with value $1$ if pixel $p$ is within the aperture and $0$ otherwise, and $I_{i,p}$ is the counts in pixel $p$ in difference image $I_i$. The covariance between aperture sums on adjacent group differences was calculated as
\begin{equation}\label{eq:covariance-between-sums}
    \text{Cov}(A_i,A_k) = \sum_p\sum_qw_{i,p}w_{k,q}\text{Cov}(I_{i,p}, I_{k,q}).
\end{equation}
In other words, the covariance between adjacent aperture sums is the total covariance across shared detector pixels between the sums, indicated by the aperture weights $w$. With these quantities, the flux $\hat{f}$ can be estimated with a generalized least squares (GLS) fit with the covariance matrix \citep[i.e., using the methods of ][]{Kubik_2015, hogg2010dataanalysisrecipesfitting}. The covariance matrix is only used for determining how each individual sum is weighted in the fit. To get an empirical measure of the noise, we define the detection {significance $\mathcal{S}$} as $\hat{f}$ compared to an ensemble of empty-source apertures $\hat{f}^{\text{null}}$ processed in the same manner: 

\begin{equation}\label{eq:detection-snr}
    {\mathcal{S}=\frac{\hat{f}-\langle\hat{f}^\text{null}\rangle}{\text{{std}}(\hat{f}^\text{null})}}.
\end{equation}

{We use $\mathcal{S}$ to quantify how strongly a recovered source is separated from the empirical null distribution. Because this statistic is normalized using a global ensemble of empty-source apertures rather than a locally estimated background, spatial variations in the image background can contribute additional scatter to $\mathcal{S}$. Consequently, $\mathcal{S}$ should not be interpreted as a photometric SNR. Photometric uncertainties were estimated independently from the scatter of the recovered source fluxes.}

\subsection{Detection catalog}

Tables \ref{tab:tycho_astrometry} and \ref{tab:tycho_gls} present 13 recovered candidate moving objects across 14 fields, of which 5 were matched to catalogued main-belt asteroids via JPL Horizons. The remaining 8 have no catalogued counterpart and are flagged as candidate new objects. Our detections span a velocity range of {$\sim4$--$37$} arcseconds per hour and a magnitude range of {$\sim21.5$--$26.8$} in both F150W2 and F322W2. We do not attempt formal orbit determination given the short observed arcs.

\begin{table*}
  \centering
  \caption{Astrometry of the PID~5594 candidate moving objects and identified main belt asteroids. Cross-matched objects are labelled with their designation, while unmatched objects are indicated with $(\cdots)$. RA/Dec and epoch are reported at a reference epoch corresponding to the first measurement. Angular speed and PA are the fitted sky-plane rate and position angle of motion (measured from North through East).}
  \label{tab:tycho_astrometry}
  \begin{tabular}{l l l r r r r}
\toprule
    Tracklet ID & Designation & {Epoch (UTC)} & RA & Dec & Speed & PA \\
     &  & & (deg) & (deg) & (arcsec/hr) & (deg) \\
    \midrule
    1 & $\cdots$ & 2024-11-28T23:48:48 & 351.87092 & -2.08811 & 22.14 & 68.9 \\
    2 & 417988 (2007 TW274) & 2024-11-30T07:14:17 & 6.56575 & 1.33644 & 6.96 & 27.6 \\
    3 & $\cdots$ & 2024-11-30T07:14:12 & 6.55187 & 1.34600 & 9.12 & 319.4 \\
    4 & $\cdots$ & 2025-06-23T21:38:01 & 181.53629 & -8.80750 & 37.26 & 108.3 \\
    5 & {(2021 MQ23)} & 2025-06-28T02:08:36 & 179.50942 & -10.75994 & 30.05 & 246.0 \\
    6 & $\cdots$ & 2025-07-05T19:26:01 & 206.88562 & -11.75289 & 12.83 & 113.2 \\
    7 & $\cdots$ & 2024-10-03T18:57:07 & 322.36925 & -7.67575 & 6.75 & 210.9 \\
    8 & $\cdots$ & 2024-10-03T18:57:15 & 322.35246 & -7.67547 & 24.14 & 34.6 \\
    9 & $\cdots$ & 2024-12-15T03:57:25 & 16.38121 & 13.39014 & 14.09 & 138.9 \\
    10 & 436248 (2010 BW65) & 2025-03-27T12:00:01 & 135.16662 & 20.90964 & 14.00 & 223.0 \\
    11 & 284091 (2005 ME4) & 2025-03-27T11:59:59 & 135.15942 & 20.88311 & 8.89 & 321.1 \\
    12 & $\cdots$ & 2025-03-27T12:00:03 & 135.15017 & 20.90236 & 7.09 & 240.0 \\
    13 & 411380 (2010 VH63) & 2025-03-27T12:13:06 & 135.15450 & 20.87864 & 4.36 & 207.2 \\
    \bottomrule
  \end{tabular}
\end{table*}

\begin{table*}
  \centering
  \small
  \setlength{\tabcolsep}{1.5pt}
  \caption{Photometry of the candidate moving objects recovered in PID~5594. Magnitudes are measured in {either a circular aperture (for sources with negligible trailing)} or a trail-aligned rectangular aperture, and have been aperture-corrected using the capture fraction measured empirically from the injection and recovery simulation (section \ref{sec:injectionRecovery}), in which sources of known flux are recovered through the identical measurement procedure. Magnitude errors are the asymmetric $1\sigma$ bounds propagated exactly from the flux $F\pm\delta F$ ($\sigma_\pm=\mp2.5\log_{10}(1\mp\delta F/F)$), where $\delta F$ is the scatter of the independent per-exposure measurements. { $\mathcal{S}$ corresponds to the detection significance and is calculated using Equation \ref{eq:detection-snr} whereas SNR refers to the photometric SNR calculated from the object's magnitude and its uncertainty.} {The values in parentheses next to the measured $\mathcal{S}$ and SNR values correspond to the factor improvement in the same quantities on the ramp-fit images.} The subscript indicates the NIRCam channel: short-wavelength (SW, F150W2) and long-wavelength (LW, F322W2).}
  \label{tab:tycho_gls}
  \begin{tabular}{l l r r r r r r}
    \toprule
    Tracklet ID & Designation & $m_{\rm AB}$(F150W2) & $m_{\rm AB}$(F322W2) & $\mathcal{S}_{\rm SW}$ & $\mathcal{S}_{\rm LW}$ & SNR$_{\rm SW}$ & SNR$_{\rm LW}$ \\
     &  & (mag) & (mag) & & & & \\
    \midrule
    1 & $\cdots$ & $25.12^{+0.09}_{-0.09}$ & $26.30^{+0.45}_{-0.32}$ & 22.7 (1.54$\times$) & 9.4 (3.03$\times$) & 12.1 (1.41$\times$) & 3.0 (1.66$\times$) \\
    2 & 417988 (2007 TW274) & $22.144^{+0.002}_{-0.002}$ & $23.00^{+0.01}_{-0.01}$ & 1198.5 (1.77$\times$) & 402.2 (2.00$\times$) & 495.8 (1.46$\times$) & 91.6 (1.10$\times$) \\
    3 & $\cdots$ & $25.95^{+0.05}_{-0.05}$ & $26.58^{+0.05}_{-0.05}$ & 27.8 (1.89$\times$) & 13.5 (2.10$\times$) & 21.7 (1.51$\times$) & 21.9 (1.80$\times$) \\
    4 & $\cdots$ & $25.08^{+0.18}_{-0.15}$ & $25.51^{+0.72}_{-0.43}$ & 18.4 (1.15$\times$) & 14.4 (1.97$\times$) & 6.7 (1.96$\times$) & 2.1 (1.20$\times$) \\
    5 & (2021 MQ23) & $23.54^{+0.03}_{-0.03}$ & $24.50^{+0.06}_{-0.06}$ & 93.2 (2.17$\times$) & 35.5 (2.79$\times$) & 33.4 (1.08$\times$) & 18.6 (1.18$\times$) \\
    6 & $\cdots$ & $25.92^{+0.05}_{-0.05}$ & $26.80^{+0.09}_{-0.08}$ & 21.1 (2.43$\times$) & 7.8 (1.92$\times$) & 20.7 (1.15$\times$) & 13.1 (1.53$\times$) \\
    7 & $\cdots$ & $23.14^{+0.01}_{-0.01}$ & $23.78^{+0.04}_{-0.04}$ & 390.6 (1.64$\times$) & 280.7 (2.39$\times$) & 125.6 (1.05$\times$) & 30.4 (1.20$\times$) \\
    8 & $\cdots$ & $24.85^{+0.04}_{-0.04}$ & $25.47^{+0.10}_{-0.09}$ & 36.6 (1.74$\times$) & 22.9 (3.42$\times$) & 27.5 (3.46$\times$) & 11.3 (1.16$\times$) \\
    9 & $\cdots$ & $24.25^{+0.27}_{-0.22}$ & $25.05^{+0.18}_{-0.16}$ & 68.6 (2.07$\times$) & 40.6 (2.73$\times$) & 4.5 (1.05$\times$) & 6.4 (1.04$\times$) \\
    10 & 436248 (2010 BW65) & $21.49^{+0.01}_{-0.01}$ & $22.33^{+0.01}_{-0.01}$ & 999.4 (1.81$\times$) & 581.6 (2.92$\times$) & 124.2 (1.62$\times$) & 73.0 (1.58$\times$) \\
    11 & 284091 (2005 ME4) & $21.60^{+0.01}_{-0.01}$ & $22.45^{+0.02}_{-0.02}$ & 1616.8 (1.89$\times$) & 645.1 (2.00$\times$) & 109.8 (1.33$\times$) & 57.8 (1.17$\times$) \\
    12 & $\cdots$ & $26.68^{+0.15}_{-0.13}$ & $25.95^{+1.17}_{-0.55}$ & 14.1 (1.98$\times$) & 27.8 (1.65$\times$) & 7.9 (1.10$\times$) & 1.5 (1.00$\times$) \\
    13 & 411380 (2010 VH63) & $22.08^{+0.03}_{-0.03}$ & $23.03^{+0.03}_{-0.03}$ & 1354.9 (1.75$\times$) & 572.4 (1.59$\times$) & 37.3 (1.04$\times$) & 34.4 (1.08$\times$)\\
    \bottomrule
  \end{tabular}
\end{table*}

\subsection{Astrometric validation of group difference images}
\begin{figure}
    \centering
    \includegraphics[width=\textwidth]{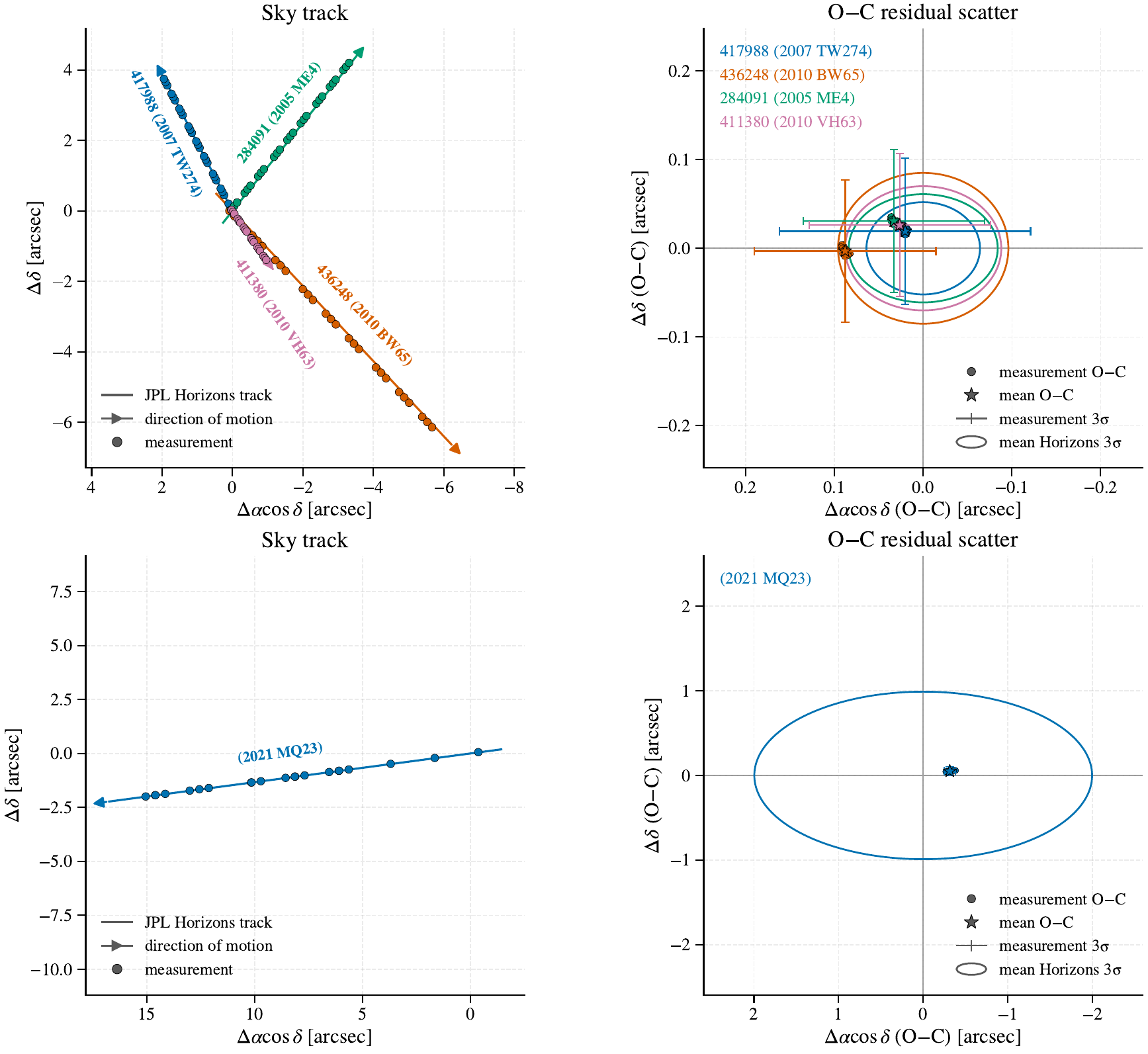}
    \caption{Astrometric validation against JPL Horizons for five main-belt asteroids: 417988 (2007 TW274), 436248 (2010 BW65), 284091 (2005 ME4), 411380 (2010 VH63) and (2021 MQ23). {Left:} sky track of the group difference detections relative to the first epoch, with the Horizons ephemeris overplotted. Arrows indicate the direction of on-sky motion. Objects are distinguished by color. {Right:} Observed-minus-computed (O$-$C) residuals, with the same color encoding. Small circles show the O$-$C residual at each individual epoch. but are too small to be clearly visible due to low astrometric scatter. Stars mark the per-object mean O$-$C. The cross on each mean point shows the 3$\sigma$ uncertainty on the mean, and the ellipse shows the 3$\sigma$ positional uncertainty of the Horizons ephemeris at the mean epoch. {The per-epoch residuals are not independent, as measurements of a given object may share common astrometric systematics that do not average down with the number of epochs. Timing uncertainties are expected to contribute negligibly at the apparent rates of the asteroids given the $\sim 0.1$ second timing precision achieved with NIRCam rolling shutter timing corrections.} All four objects show O$-$C residuals consistent with their predicted position from Horizons.}

    \label{fig:asteroid-417988-obs}
\end{figure}

We serendipitously recovered five known asteroids (Table~\ref{tab:tycho_gls}) in this data set, {first identified} by crossmatching against the \href{https://ssd.jpl.nasa.gov/tools/sbdb_lookup.html#/}{JPL Horizons small body identifier} {and later verified through the Minor Planet Center (MPC)'s identification process after submission of the astrometry}. We validate the astrometric accuracy of our calibration procedure's finer temporal sampling by comparing the observed motion of these asteroids with their predicted ephemeris in Figure \ref{fig:asteroid-417988-obs}. {Astrometric measurements, as well as the corresponding positional and timing uncertainties, were acquired by following the procedures outlined for JWST/NIRCam in \cite{sarc-manual}. We refined the image WCS using Gaia Data Release 3 (DR3) \citep[][]{gaiaDR3}. The effective epoch of each group-difference measurement was calculated from the detector readout timing. Timing offsets were applied based on the individual asteroid's position on the specific detector submodule to correct for NIRCam's $\sim 11$ second rolling shutter \citep[][]{jdox2016}.} The measured astrometric positions of the five asteroids are consistent with their predicted positions from Horizons, with O$-$C residuals within the $3\sigma$ ephemeris uncertainty for all objects, confirming that the group difference images are suitable for precise astrometry and motion recovery.

{The residual O$-$C displacements in Figure \ref{fig:asteroid-417988-obs} could arise from both systematic offsets in the WCS solution and systematic errors in the assigned observation times. A timing error would produce an O$-$C displacement parallel or antiparallel to the angular velocity vector, with $\Delta\theta_{\rm time}\simeq{\mu}\Delta t$. When applying the row-level timing corrections outlined in \cite{sarc-manual}, conservative timing uncertainties of $\sim0.1$ seconds can be achieved. Given this timing precision and the relatively modest angular velocities of the recovered asteroids, the expected astrometric contribution from timing uncertainty is substantially smaller than the measured residuals. In addition, the residual vectors are not consistently aligned with the angular velocity vectors. We therefore interpret the remaining offsets as being primarily associated with the astrometric WCS solution and ephemeris uncertainty.}

\subsection{Faint discovery}
\begin{figure}
    \centering
    \includegraphics[width=\textwidth]{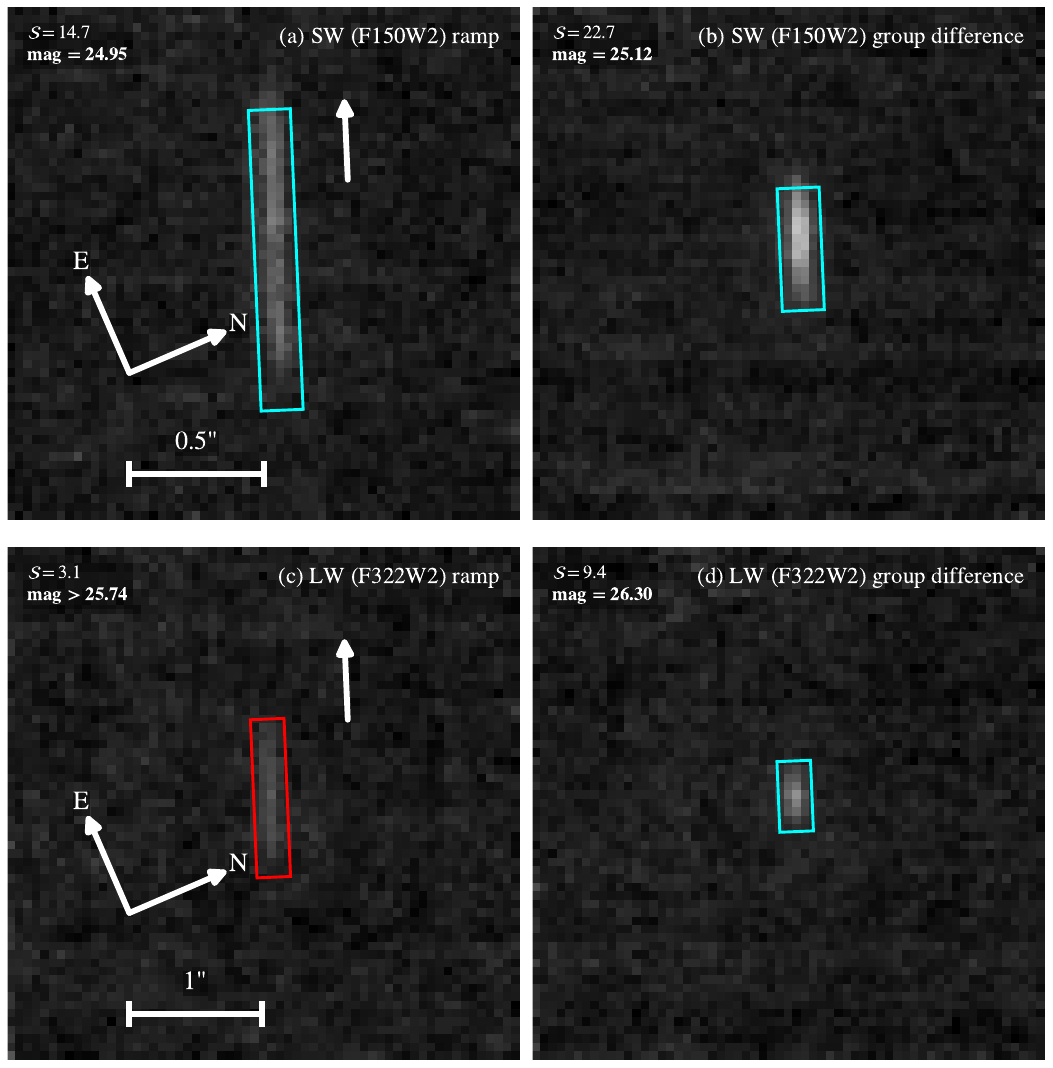}
    \caption{Faint-limit recovery of candidate moving object Tracklet 1. (a) A stack of $9$ short-wavelength ramp-fit exposures. The individual exposures making up the stack of images have effective integration times of $T_\text{int}=204$ seconds. The asteroid appears as an elongated trail. (b) $27$ shift-and-stacked group differences, with the same total exposure time but with the individual images comprising the stack having an effective exposure of $51$ seconds. The asteroid's trail is shortened by a factor of $\sim$2. (c-d) The same as (a-b), but in the long-wavelength detector, which has a $2\times$ coarser pixel scale. {${\mathcal{S}}$ is calculated using Equation \ref{eq:detection-snr}.} The object falls below our detection threshold of ${\mathcal{S}}=5$ in the ramp-fit image in F322W2. It is recovered in the corresponding group-difference image. We define a channel with ${\mathcal{S}}<5.0$ (Equation \ref{eq:detection-snr}) as a non-detection, and for non-detected channels we report a $5\sigma$ upper limit on the source flux.}
    \label{fig:faint-discovery}
\end{figure}

Group differencing can recover moving objects that are below the detection threshold in the standard ramp-fit product. An example, Tracklet 1 (first row of Tables \ref{tab:tycho_astrometry} and \ref{tab:tycho_gls}), is shown in Figure \ref{fig:faint-discovery}. It is a previously unidentified mover whose ramp-fit {detection significance} fell below our adopted detection threshold of ${\mathcal{S}=5}$ in F322W2. In the corresponding group-differenced image, the object is recovered at ${\mathcal{S}}\approx9.4$, demonstrating that working with the group-difference images can extend detectability into a regime that the standard pipeline products cannot access. 

\subsection{Flux conservation and SNR improvements}
\begin{figure}
    \centering
    \includegraphics[width=\textwidth]{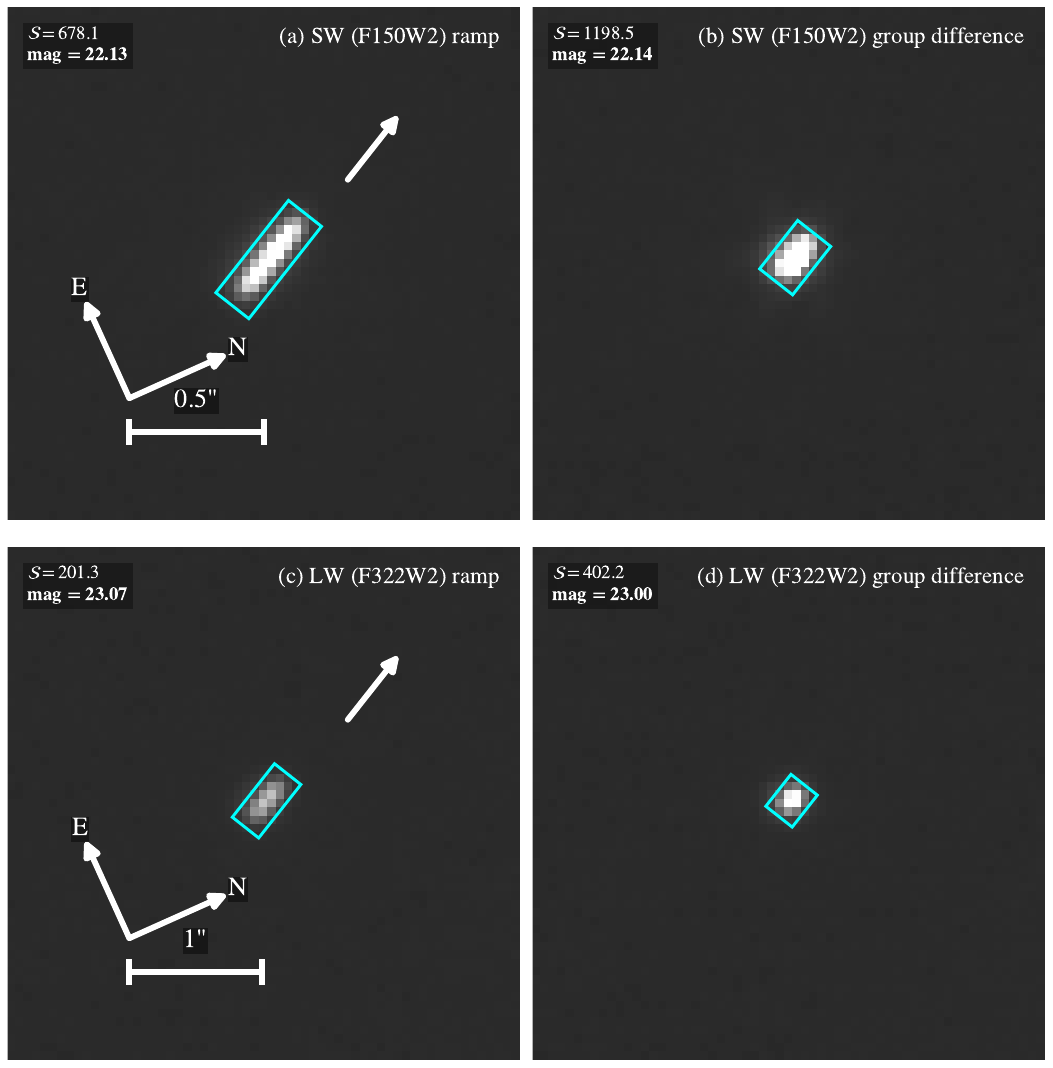}
    \caption{Shift-and-stacked exposures of asteroid 417988 serendipitously recovered in PID 5594 imagery, shown in the same panel layout as Figure \ref{fig:faint-discovery}.}
    \label{fig:ramp-vs-group-diff}
\end{figure}

The SNR improvement of asteroid 417988 is shown in Figure \ref{fig:ramp-vs-group-diff}, shown in the same panel layout as Figure \ref{fig:faint-discovery}. Group differencing improves the detectability of 417988 against the background when compared to the full ramp fit by a factor of $\sim1.7$--$2$.

The GLS estimator is unbiased for any choice of weights, provided the assumed linear ramp model (constant flux rate) correctly describes the data. For a source that only briefly transits the pixel, the constant-flux model is misspecified. We showed in Section \ref{sec:nircam-structure} that in the Poisson-noise limit, the pipeline's weights reduce the flux estimate to be the difference between the last group and the first group in the integration, divided by the total integration time. This is independent of which group(s) the source actually illuminated the pixel, and holds for each pixel along the trail.

In the read-noise limit, the pipeline's slope fit reduces to an ordinary-least-squares (OLS) fit to the reads. OLS weights each group in proportion to its time offset from the ramp midpoint, $t_i-\bar t$, so groups near the start and end of the ramp contribute more to $\hat{f}$ than groups near the middle. For a transiting source, the per-pixel $\hat{f}$ therefore depends on \textit{when} during the ramp the flux arrived, and individual pixels along a trail can be biased relative to the true instantaneous rate, as the object moves during the exposure. However, because these weights are symmetrical, summing over the trail spanning the full ramp will cancel out this position-dependent bias exactly, so the total flux is conserved regardless of the group-noise regime. 
We can therefore compare against this flux sum to check whether we correctly recover the total flux with our group differencing method.

\begin{figure}
    \centering
    \includegraphics[width=\textwidth]{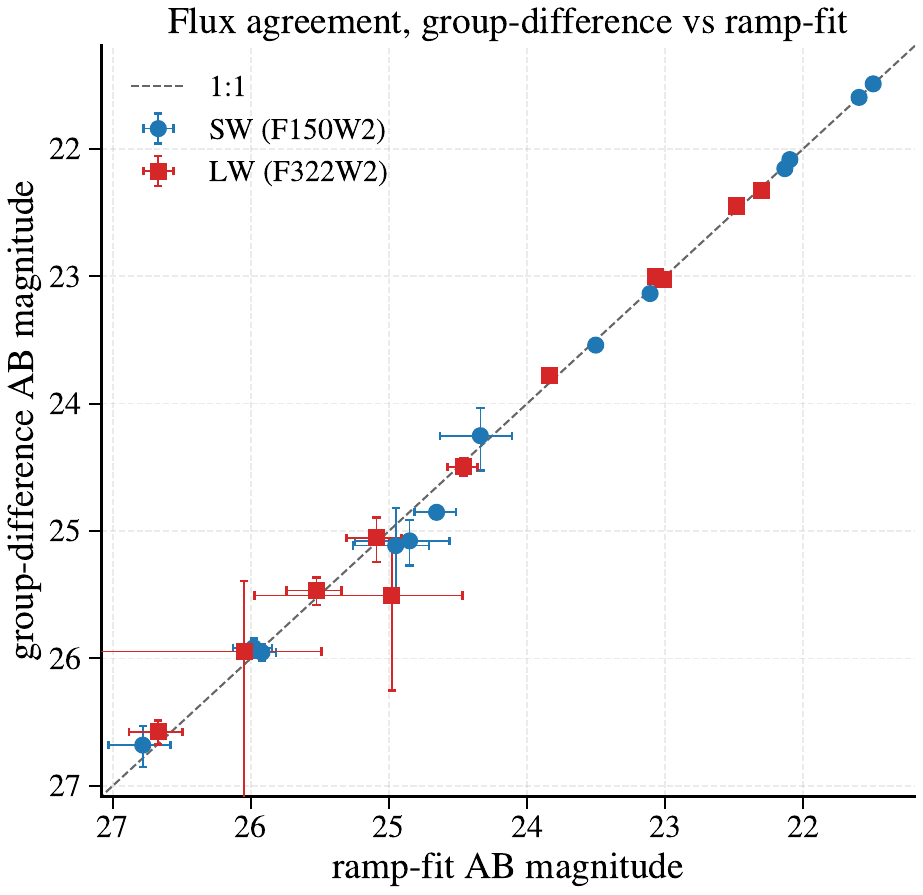}
    \caption{Flux agreement between ramp-fit and group-difference estimates for our candidate moving sources (See Tables \ref{tab:tycho_astrometry} and \ref{tab:tycho_gls}). Only sources with $\text{SNR}>5$ in both the ramp-fit and group-differenced images are plotted here. All magnitudes agree within their uncertainties, demonstrating that flux is conserved across both image calibrations.} 
    \label{fig:flux-agreement}
\end{figure}

For group differencing, each image is by construction the total counts accumulated within the specific group interval, so a GLS fit to these images is unbiased in all cases. Figure \ref{fig:flux-agreement} confirms this empirically by plotting the measured magnitude of our candidate movers in both the group-difference and ramp-fit images, showing that they agree to within photometric uncertainty.

\subsection{Injection/Recovery Simulation}\label{sec:injectionRecovery}
To estimate the expected gain in sensitivity from using group differencing to recover trailed moving objects, we use an injection/recovery simulation to evaluate the calibration procedure's ability to extend survey sensitivity to fainter trailed sources in the background-noise limited regime. Synthetic point sources were injected at the ramp level, prior to any calibration, so that both the \texttt{calints.fits} and \texttt{calgroups.fits} products are derived from the same injected signal. For each injection, we model the source as a WebbPSF \citep{webbPSF} point-spread function swept along a linear trajectory at the assigned apparent rate of motion. Poisson noise is added based on the expected source flux, converted from its AB magnitude in the F150W2/F322W2 filters into detector electrons using the image calibration information. The PSF is integrated over each frame time, and the resulting per-group charge is added cumulatively to the up-the-ramp signal of the overlapping pixels such that the injected source trails realistically within the integration. A small fraction of faint injected sources are recovered with fluxes substantially exceeding their injected values, owing to contamination from static field sources. These sources are classified as contaminated detections rather than successful recoveries, and counted against the completeness fraction. An empirical contamination threshold was defined from the recovered-to-injected flux ratio distribution of bright ($m_{\rm AB}<24$) sources, for which contamination is negligible. As contamination could only bias the flux ratio upwards, the expected flux ratio was taken as the median of the lower 50\% of this distribution. Detections whose recovered-to-injected flux ratio exceeded this value by more than three times its scatter were rejected. The same threshold was applied uniformly to both the ramp-fit and group-difference reductions. The completeness $\mathcal{C}$ in each grid cell is the recovered fraction. With 30 trials per cell, the binomial uncertainty on $\mathcal{C}$ is at most $9\%$.

\begin{figure}
    \centering
    \includegraphics[width=\textwidth]{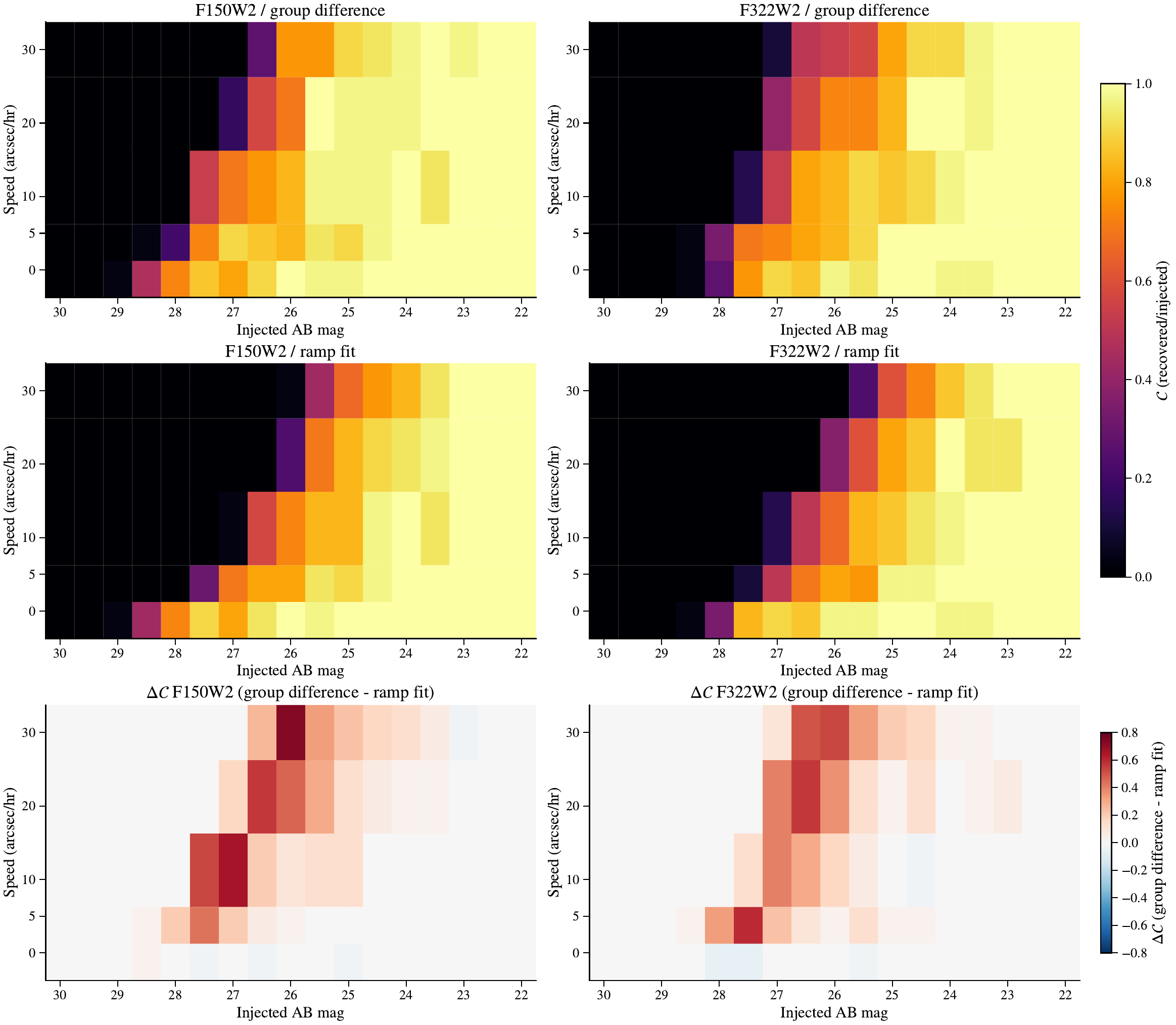}
    \caption{Injection-recovery completeness $\mathcal{C}$ (fraction recovered/injected) for synthetic point sources injected into the raw ramps of PID 5594 and reduced through both pipelines, as a function of injected AB magnitude and apparent speed. Columns compare the group difference reductions (top) against the standard ramp-fit products (middle) for F150W2 (left) and F322W2 (right). {The difference in completeness ($\Delta \mathcal{C}$) between the group difference and ramp-fit products are shown on the bottom}. At injected speeds $v$ higher than $5\arcsec/\text{hr}$, group difference reductions reach fainter 50\% completeness limits, with the gain increasing toward higher apparent rates. At $v=0$, both reduction methods achieve the same sensitivity.}
\label{fig:pointSourceInjectionRecovery}
\end{figure}

\begin{figure}
    \centering
    \includegraphics[width=\textwidth]{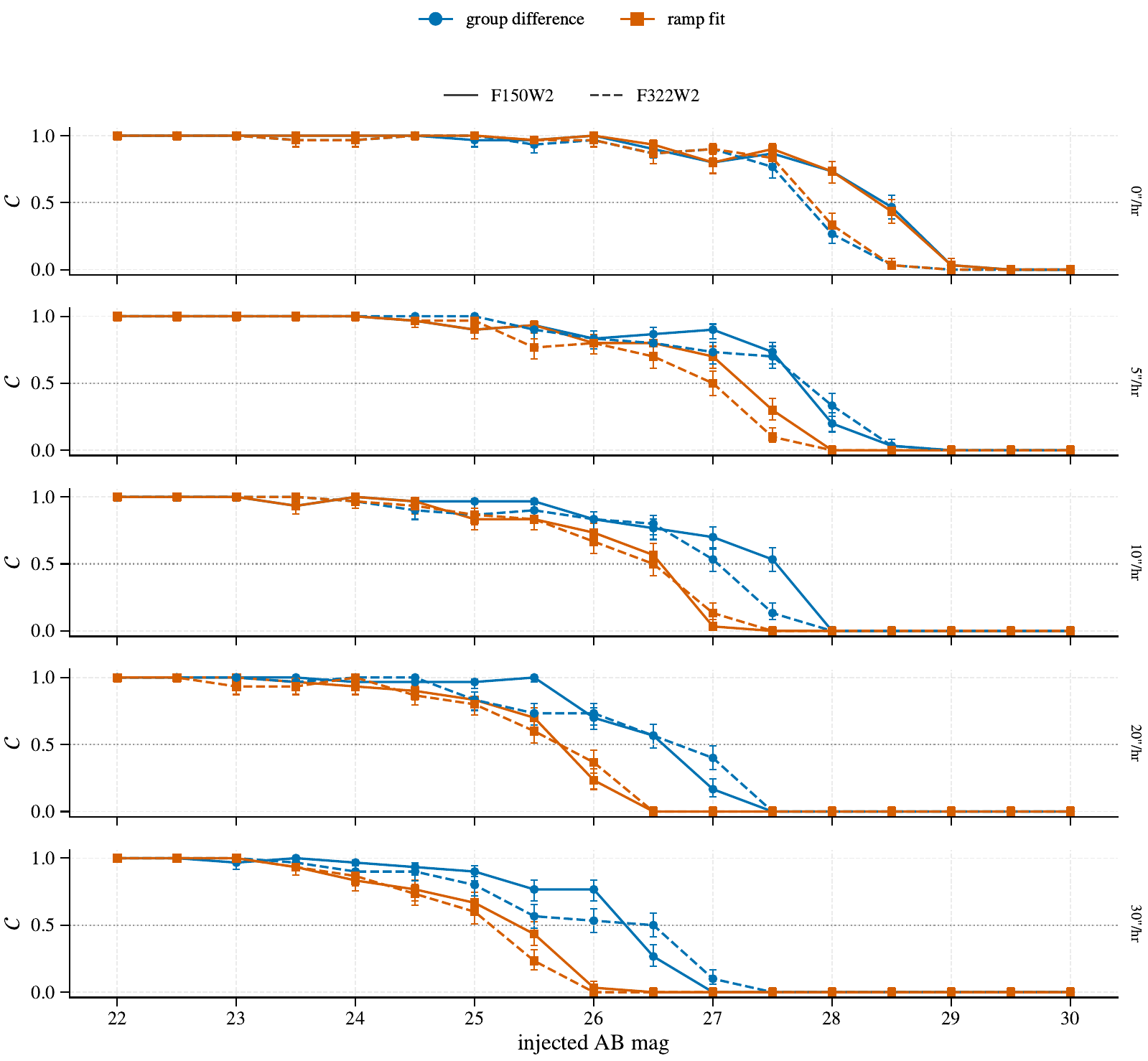}
    \caption{Injection-recovery completeness $\mathcal{C}$ as a function of injected AB magnitude, shown separately for each angular rate of motion $v$ (0, 5, 10, 20, 30 arcsec/hr; one row per rate). In each panel, blue circles show the group-difference calibration and orange squares show the standard ramp-fit pipeline; solid lines are F150W2 and dashed lines are F322W2. Error bars are binomial confidence intervals on the recovered fraction at each magnitude/rate/filter cell (30 injected sources per cell). The dotted horizontal line marks 50\% completeness. We denote the magnitude at which a given curve crosses this line as $m_{\rm 50\%}$. At zero apparent rate, the two calibration branches are statistically indistinguishable ($m_{\rm 50\%}$ differs by $\leq0.05$ mag in both filters), as expected since group-differencing and standard ramp-fitting are mathematically equivalent for a stationary source. As the angular rate increases, group-difference calibration recovers systematically fainter sources than the ramp-fit pipeline, with the increase in $m_{\rm 50\%}$ growing from $\sim$0.4–0.7 mag at 5 arcsec/hr to $\sim$0.9–1.3 mag at 30 arcsec/hr in both filters.}
    \label{fig:completeness-curves}
\end{figure}

\begin{figure}
    \centering
    \includegraphics[width=\textwidth]{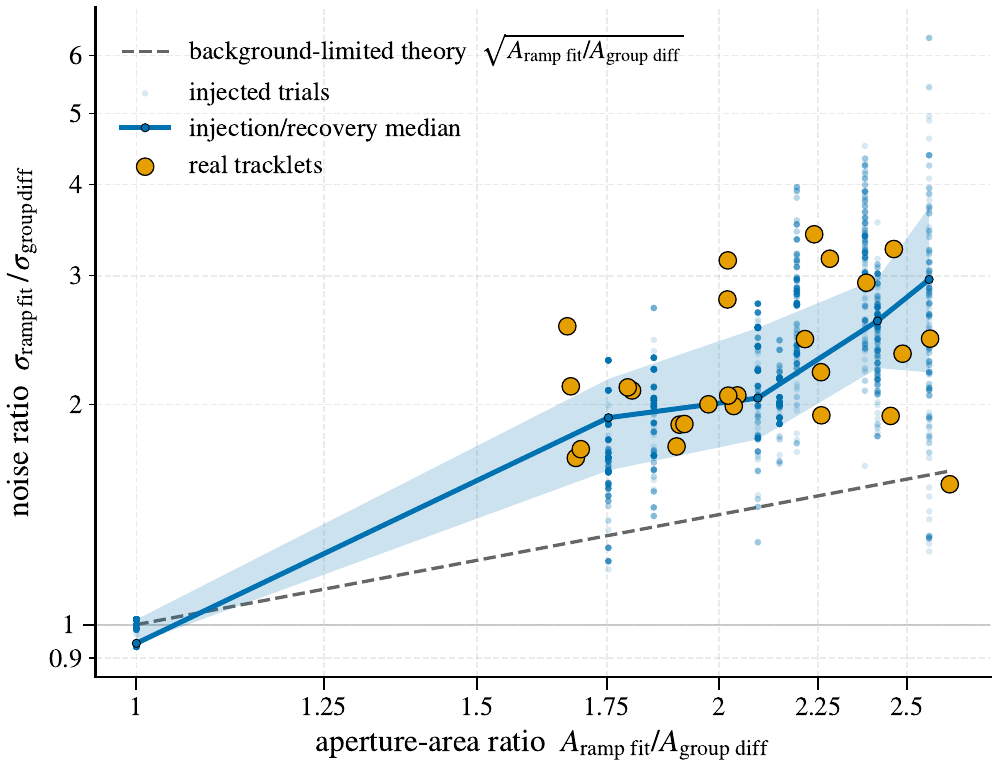}
    \caption{Detection-noise ratio $\sigma_{\rm ramp\ fit}/\sigma_{\rm group\ diff}$ vs. aperture-area ratio $A_{\rm ramp \ fit}/A_{\rm group \ diff}$. Both F150W2 and F322W2 measurements have been aggregated together. Small points are individual injected trials. The solid line and shaded band give the per-speed median and $1\sigma$. As the chosen apertures in ramp-fit and group-difference images nominally measure the same source flux, the SNR boost is driven by the reduction in the noise in the group-difference aperture resulting from its smaller size. The filled orange circles are the real candidate mover measurements. At $v=0$ (aperture area ratio of 1), the noise ratio is close to unity, with a median value of $0.94\pm0.01$. At non-zero speeds, the noise in each aperture grows beyond that expected from a uniform white background (i.e., $\sigma\propto\sqrt{A}\propto\sqrt{n_\text{pix}}$, Equation \ref{eq:snr}). This is likely caused by additional sources of systematic noise in each moving aperture, such as PSF wing contamination, a spatially varying background, detector systematics, etc. We do not attempt to model systematic noise explicitly, and instead characterize it empirically within our simulation. The noise ratio for our real candidate movers is consistent with our injected sources in our simulation.}
    \label{fig:SNR-boost-comparison}
\end{figure}

Figure \ref{fig:pointSourceInjectionRecovery} shows the resulting completeness maps, and Figure \ref{fig:completeness-curves} shows the completeness curves at each speed and filter. At $v=0$, both methods achieve the same limiting magnitude of $\sim28.5$ AB mag at ${\mathcal{S}}=5$ in both F150W2 and F322W2, which is consistent with the quoted $m_\text{lim}=29$ AB mag in both filters at ${\mathcal{S}}=3$ for PID 5594 \citep[][]{PIDSURVEYSLICE}. At angular speeds $v>5$ arcsec/hr, the group difference reductions reach fainter $50\%$ completeness limits than the standard ramp-fit products, with the gain increasing towards higher rates. At the highest injected rate of 30 arcsec/hr, the 50\% completeness limit from group differencing improves by $\sim0.9$--$1.3$ mag in either filter, above the $2.5\log_{10}(\sqrt{N-1})=0.6$ mag floor set by the background-limited case for the $N=4$ group exposures PID 5594. Because each group-difference image spans a shorter exposure than the full $N$-group ramp, the moving object trail in a group-difference image is reduced such that the photometric aperture with area $A$ sized to enclose it is smaller than the aperture required in the standard ramp-fit product ($A_{\rm ramp \ fit}\geq A_{\rm group \ diff}$).  Figure \ref{fig:SNR-boost-comparison} plots the resulting noise ratio $\sigma_{\rm ramp\ fit}/\sigma_{\rm group\ diff}$ as a function of the aperture-area ratio $A_{\rm ramp\ fit}/A_{\rm group \ diff}$, where $\sigma$ is the empirical aperture noise measured from an ensemble of empty-source moving apertures (denominator of Equation \ref{eq:detection-snr}). The measured noise ratio exceeds the $\sqrt{A_{\rm ramp \ fit}/A_{\rm group \ diff}}$ prediction for statistically independent pixels. This scaling is expected to be pipeline-dependent, and discussed further in Section \ref{sec:discussion}.

\begin{figure}
    \centering
    \includegraphics[width=\textwidth]{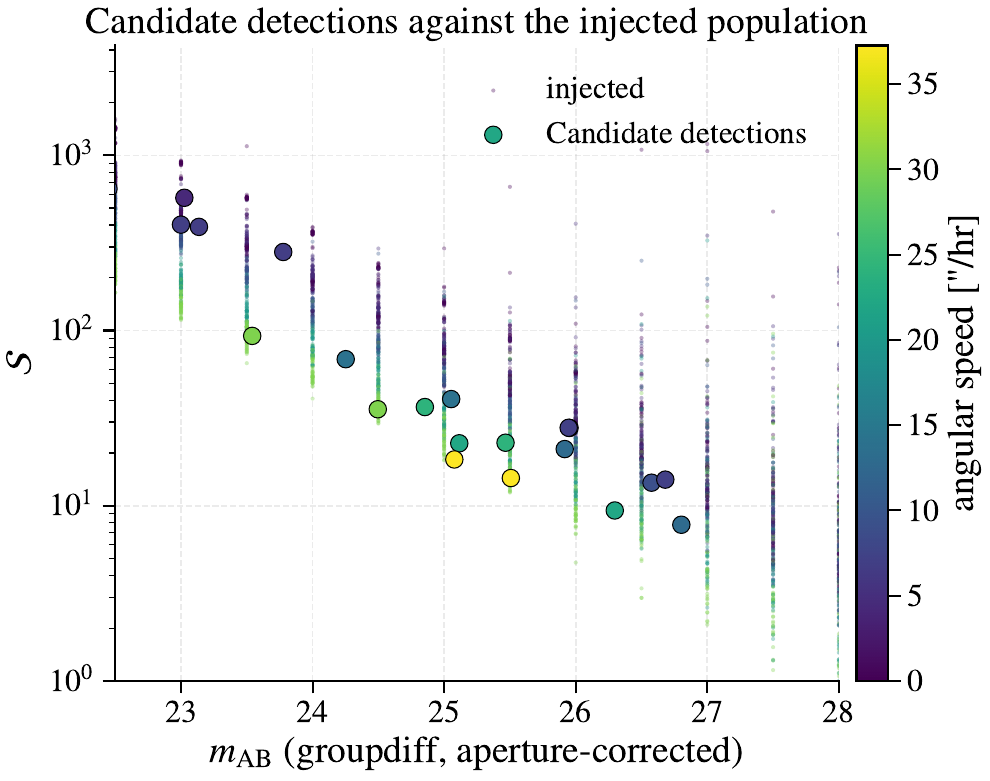}
    \caption{Detection Significance (Equation \ref{eq:detection-snr}) vs.\ apparent AB magnitude for injected point sources and our 13 candidate movers. All measurements were made on the group-difference images. The candidate movers fall within the scatter of the injected SNR--magnitude relationship, confirming that the synthetic sources reproduce expected SNR behaviour of real detections and validates their use in the completeness analysis of Figure \ref{fig:pointSourceInjectionRecovery}.}
    \label{fig:mag-vs-snr}
\end{figure}

In order to verify that the injected source population is a realistic proxy for the real moving objects, we plot $\mathcal{S}$ vs.\ magnitude for all sources in Figure \ref{fig:mag-vs-snr}. The injected sources follow the same trend as the real recovered moving objects, demonstrating consistency between the injected sources and the real candidate moving sources.

\section{Discussion: Improvement in Sensitivity and Limitations}
\label{sec:discussion}

Figures \ref{fig:pointSourceInjectionRecovery} and \ref{fig:completeness-curves} show a sensitivity increase as large as $1.3$ AB mag between group differencing and ramp-fitting for our dataset, around 1 AB mag deeper than predicted by a background-limited regime. As shown in Figure \ref{fig:SNR-boost-comparison}, this excess arises because the smaller aperture used for group-difference images reduces the aperture noise by more than the $\sqrt{A_{\rm ramp \ fit}/A_{\rm group \ diff}}$ scaling expected for independent, uncorrelated pixels. The excess noise in each aperture can be attributed to a variety of sources that do not average down as independent, uncorrelated noise, such as spatially varying background/detector systematics, contamination from the PSF wings of imperfectly masked stationary sources, as well as unresolved source confusion. These are generic features of NIRCam imaging, such that the excess noise we measure here is plausibly representative of typical archival NIRCam pointings. As moving sources traverse the detector, these noise sources can enter the aperture and will not average down simply. This gain has dependence on both the field itself (crowdedness), the trajectory of the source, as well as the calibration/detection pipeline. For example, different-sized apertures, more robust stationary source masking and detector systematic removal can affect the aperture noise scaling. We therefore maintain the white-noise background-limited prediction as a minimum expectation for the sensitivity increase of group-differencing for moving sources, and interpret our highest measured gain of $0.9$--$1.3$ mag as an empirical value for this and similar datasets.

\begin{figure}
    \centering
    \includegraphics[width=\textwidth]{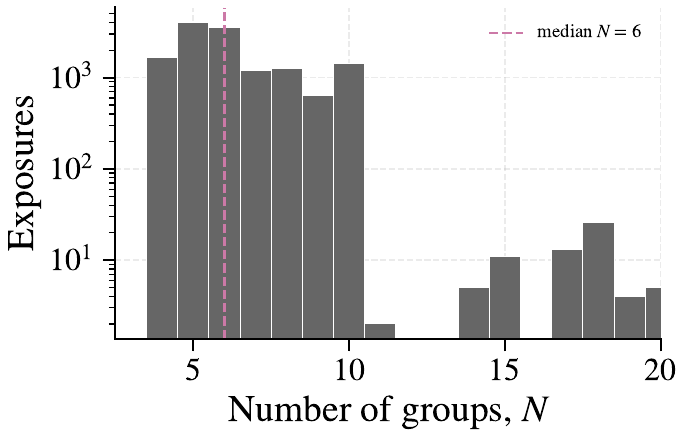}
    \caption{Distribution of all available NIRCam imaging, $|\beta|<10^\circ$, as of September 2026. The median number of groups per integration $N=6$ (dashed line) corresponds to a minimum sensitivity gain for trailed sources of 0.9 mag if reprocessed with group differencing, under the background-limited noise floor. Scaling our empirically measured average sensitivity increase of $1.1$ mag for trailed sources at $N=4$ groups (from injection/recovery simulations) by the same $\sqrt{N-1}$ SNR dependence yields an estimated gain of 1.4 mag at $N=6$.}
    \label{fig:archival-census}
\end{figure}

The JWST/NIRCam archive contains thousands of imaging exposures near the ecliptic spanning a broad range of filters, depths, and readout patterns. Although the heterogeneous nature of the archive prevents a simple global completeness estimate, these data represent a substantial untapped resource for serendipitous solar system science. Figure \ref{fig:archival-census} shows a histogram of the number of exposures vs.\ the number of groups $N$. There are 13,995 \texttt{*uncal.fits} exposures within $10^\circ$ of the ecliptic, with a median integration length of $N_{\rm med}=6$. In the background-noise limit, shift-and-stacking the $N-1$ resultant group differences along the source's trajectory using the covariance matrix recovers an SNR gain of $\sqrt{N_{\rm med}-1}=\sqrt{5}\approx2.2$ over the trailed full-ramp detection, corresponding to a minimum limiting-depth increase of $\Delta m=2.5\log_{10}\sqrt{N_{\rm med}-1}\approx 0.9$ mag. Adopting the average value of our empirically measured sensitivity increase of $1.1$ mag at $N=4$ groups from our injection/recovery simulations, and assuming the increase scales similarly, the estimated limiting depth increase can reach values of $1.1\text{ mag}+2.5\log_{10}(\sqrt{\frac{5}{3}}) =1.4\,\text{ mag}$.


Group differencing does not provide any advantages over the standard ramp-fit for objects that do not demonstrate any intra-exposure trailing. Therefore, for all but the longest integrations, the method presented in this paper is better suited for objects with high angular velocities on the sky, such as near-Earth asteroids {(NEAs)}. {In the NEA regime, accurate observation timing becomes increasingly important because timing errors translate directly into astrometric offsets along the angular velocity vector. Appropriate JWST/NIRCam timing procedures are described in \cite{sarc-manual}.}

Our use of Tycho Tracker \citep[][]{parrottTychoTrackerNew} directly on the group difference images has several limitations. Firstly, Tycho Tracker does not co-add our group difference images according to their covariance matrix. For trailed sources, this is not an issue, as they land on different detector pixels in each group difference frame and thus are approximately independent in time. For sources whose PSFs overlap between group difference images, their variance will be overestimated. Secondly, we expect trailing losses to diminish our sensitivity, as Tycho Tracker uses a compact PSF rather than a streak-matched filter to detect sources after stacking. Our injection/recovery simulation, which simulates trailing sources and uses rectangular apertures aligned to the sources, does not share these limitations and is a realistic estimate of the expected sensitivity increase from group differencing.

A limitation of the present simulation is that it combines the images along the a priori known trajectories of the injected sources. The optimal source detector for JWST imagery would be a blind shift-and-stack algorithm that co-adds the images using their covariance matrix as well as accommodates trailed sources by using an elongated aperture/matched-filter, but remains beyond the scope of this work.

The methods presented in this work were only applied to NIRCam. In principle, they could be applied to any up-the-ramp imager, such as other instruments on JWST or instruments on future observatories (e.g., with the Roman Space Telescope).

\section{Conclusions}\label{sec:conclusion}
This work demonstrates that differences of successive group reads within JWST NIRCam integrations enable robust and higher-SNR detection of rapidly moving solar system objects in archival imaging data. The presented technique recovers the intra-exposure temporal information within single JWST integrations. By operating on calibrated group differences rather than final ramp-fit images, the source flux can be extracted only from the inter-group intervals in which each detector pixel receives source signal, avoiding the accumulation of additional background noise during intervals when that pixel is not illuminated by the moving source and improving SNR. The approach can be used to enable both more complete searches and more precise flux measurements of rapidly moving faint solar system objects in JWST/NIRCam or in any other up-the-ramp imager.

\begin{acknowledgments}
This work is based on observations made with the NASA/ESA/CSA James Webb Space Telescope, obtained from the Mikulski Archive for Space Telescopes at the Space Telescope Science Institute. Specific datasets are referenced using their respective Digital Object Identifiers (DOIs). These observations are associated with JWST program 5594. A.G. acknowledges support from the Natural Sciences and Engineering Research Council of Canada Ontario Graduate Scholarship. This research made use of the JPL Horizons ephemeris service and NASA's Astrophysics Data System. We acknowledge the use of AI-assisted tools (Claude Code) in the preparation of this manuscript. All scientific interpretations, results, discussions, and conclusions are the responsibility of the authors. The code used to create calibrated JWST differenced images is available at the author's GitHub repository: \href{https://github.com/Corrado777/JWST-Ramp-Slicer}{https://github.com/Corrado777/JWST-Ramp-Slicer}; {Version 1.1.1 has been deposited to Zenodo: \dataset[https://doi.org/10.5281/zenodo.22664430]{https://doi.org/10.5281/zenodo.22664430}}
\end{acknowledgments}

\begin{contribution}
A.G. conceived the project, developed the methodology, performed the analysis, produced the figures, interpreted the results, and wrote the original manuscript draft. S.M. supervised the project, provided scientific guidance, reviewed the manuscript, and contributed to revisions. Both authors approved the final manuscript. 
\end{contribution}

%
\facilities{James Webb Space Telescope}

\software{astropy \citep{2013A&A...558A..33A,2018AJ....156..123A,2022ApJ...935..167A},          Source Extractor \citep{1996A&AS..117..393B}, SEP \citep{SEP-library}, jwst \citep{bushouse_2026_20058613}, WebbPSF \citep{webbPSF}, Claude Code \citep{claude}}


%


\bibliography{groupDifferencingEnhancesTheDetectabilityOfAsteroidsInJWSTNIRCamImagery}{}

@software{claude,
  author    = {{Anthropic}},
  title     = {Claude Code},
  year      = {2026},
  publisher = {Anthropic},
  url       = {https://code.claude.com},
  note      = {Agentic coding assistant. Multiple Claude models (including Sonnet and Opus) were used during software development. Accessed 2026-08-07.}
}

@article{burdanovJWSTSightingDecametre2025,
	title = {{JWST} sighting of decametre main-belt asteroids and view on meteorite sources},
	volume = {638},
	copyright = {2024 The Author(s), under exclusive licence to Springer Nature Limited},
	issn = {1476-4687},
	url = {https://www.nature.com/articles/s41586-024-08480-z},
	doi = {10.1038/s41586-024-08480-z},
	language = {en},
	number = {8049},
	urldate = {2025-10-18},
	journal = {Nature},
	publisher = {Nature Publishing Group},
	author = {Burdanov, Artem Y. and de Wit, Julien and Brož, Miroslav and Müller, Thomas G. and Hoffmann, Tobias and Ferrais, Marin and Micheli, Marco and Jehin, Emmanuel and Parrott, Daniel and Hasler, Samantha N. and Binzel, Richard P. and Ducrot, Elsa and Kreidberg, Laura and Gillon, Michaël and Greene, Thomas P. and Grundy, Will M. and Kareta, Theodore and Lagage, Pierre-Olivier and Moskovitz, Nicholas and Thirouin, Audrey and Thomas, Cristina A. and Zieba, Sebastian},
	month = feb,
	year = {2025},
	pages = {74--78},
}

@misc{hogg2010dataanalysisrecipesfitting,
      title={Data analysis recipes: Fitting a model to data}, 
      author={David W. Hogg and Jo Bovy and Dustin Lang},
      year={2010},
      eprint={1008.4686},
      archivePrefix={arXiv},
      primaryClass={astro-ph.IM},
      url={https://arxiv.org/abs/1008.4686}, 
}

@ARTICLE{parrottTychoTrackerNew,
       author = {{Parrott}, D.},
        title = "{Tycho Tracker: A New Tool to Facilitate the Discovery and Recovery of Asteroids Using Synthetic Tracking and Modern GPU Hardware (Abstract)}",
      journal = {\jaavso},
         year = 2020,
        month = dec,
       volume = {48},
       number = {2},
        pages = {262},
       adsurl = {https://ui.adsabs.harvard.edu/abs/2020JAVSO..48..262P}
}

@article{heinzeDigitalTrackingObservations2015,
	title = {Digital {Tracking} {Observations} {Can} {Discover} {Asteroids} {Ten} {Times} {Fainter} than {Conventional} {Searches}},
	volume = {150},
	issn = {1538-3881},
	url = {http://arxiv.org/abs/1508.01599},
	doi = {10.1088/0004-6256/150/4/125},
	number = {4},
	urldate = {2025-09-21},
	journal = {The Astronomical Journal},
	author = {Heinze, Aren and Metchev, Stanimir and Trollo, Joseph},
	month = sep,
	year = {2015},
	note = {arXiv: 1508.01599 [astro-ph]},
	pages = {125},
}

@article{burdanovGPUbasedFrameworkDetecting2023,
	title = {{GPU}-based framework for detecting small {Solar} system bodies in targeted exoplanet surveys},
	volume = {521},
	issn = {0035-8711},
	url = {https://doi.org/10.1093/mnras/stad808},
	doi = {10.1093/mnras/stad808},
	number = {3},
	urldate = {2025-04-26},
	journal = {Monthly Notices of the Royal Astronomical Society},
	author = {Burdanov, A Y and Hasler, S N and de Wit, J},
	month = may,
	year = {2023},
	pages = {4568--4578},
}

@misc{RampsEnablingSolar2026,
       author = {{Golovich}, Nathan Ryan},
        title = "{Up and Across the Ramps: Enabling Solar System Science from Roman Level One Data}",
 howpublished = {NASA Proposal ID. 24-ROMAN24-0095},
         year = 2024,
        month = jan,
        pages = {95},
       adsurl = {https://ui.adsabs.harvard.edu/abs/2024ngr..prop...95G}
}

@article{mullerAsteroidsSeenJWSTMIRI2023,
	title = {Asteroids seen by {JWST}-{MIRI}: {Radiometric} {Size}, {Distance} and {Orbit} {Constraints}},
	volume = {670},
	issn = {0004-6361, 1432-0746},
	shorttitle = {Asteroids seen by {JWST}-{MIRI}},
	url = {http://arxiv.org/abs/2302.06921},
	doi = {10.1051/0004-6361/202245304},
	urldate = {2025-12-15},
	journal = {Astronomy \& Astrophysics},
	author = {Müller, T. G. and Micheli, M. and Santana-Ros, T. and Bartczak, P. and Oszkiewicz, D. and Kruk, S.},
	month = feb,
	year = {2023},
	note = {arXiv:2302.06921 [astro-ph]},
	pages = {A53},
}

@article{Kubik_2015,
   title={Optimization of the multiple sampling and signal extraction in nondestructive exposures},
   volume={1},
   ISSN={2329-4124},
   url={http://dx.doi.org/10.1117/1.JATIS.1.3.038001},
   DOI={10.1117/1.jatis.1.3.038001},
   number={3},
   journal={Journal of Astronomical Telescopes, Instruments, and Systems},
   publisher={SPIE-Intl Soc Optical Eng},
   author={Kubik, Bogna and Barbier, Remi and Castera, Alain and Chabanat, Eric and Ferriol, Sylvain and Smadja, Gerard},
   year={2015},
   month=July, pages={038001} }

@ARTICLE{SEP-library,
       author = {{Barbary}, Kyle},
        title = "{SEP: Source Extractor as a library}",
      journal = {The Journal of Open Source Software},
         year = 2016,
        month = oct,
       volume = {1},
       number = {6},
          eid = {58},
        pages = {58},
          doi = {10.21105/joss.00058},
       adsurl = {https://ui.adsabs.harvard.edu/abs/2016JOSS....1...58B}
}

@MISC{jdox2016,
        key = {STScI},
        author = {{STScI}},
        title = "{JWST User Documentation (JDox)}",
 howpublished = {JWST User Documentation Website},
         year = 2016,
        month = jan,
       adsurl = {https://ui.adsabs.harvard.edu/abs/2016jdox.rept......}
}

@MISC{bushouse_2026_20058613,
  author       = {Bushouse, Howard and
                  Eisenhamer, Jonathan and
                  Dencheva, Nadia and
                  Davies, James and
                  Greenfield, Perry and
                  Morrison, Jane and
                  Hodge, Phil and
                  Simon, Bernie and
                  Grumm, David and
                  Droettboom, Michael and
                  Slavich, Edward and
                  Sosey, Megan and
                  Pauly, Tyler and
                  Miller, Todd and
                  Jedrzejewski, Robert and
                  Hack, Warren and
                  Davis, David and
                  Crawford, Steven and
                  Law, David and
                  Gordon, Karl and
                  Regan, Michael and
                  Cara, Mihai and
                  MacDonald, Ken and
                  Bradley, Larry and
                  Shanahan, Clare and
                  Jamieson, William and
                  Teodoro, Mairan and
                  Williams, Thomas and
                  Pena-Guerrero, Maria and
                  Graham, Brett and
                  Molter, Edward and
                  Brandt, Timothy and
                  Hayes, Christian and
                  Cooper, Rachel and
                  Clarke, Melanie and
                  Filippazzo, Joseph},
  title        = {JWST Calibration Pipeline},
  month        = may,
  year         = 2026,
  publisher    = {Zenodo},
  version      = {2.0.1},
  doi          = {10.5281/zenodo.20058613},
  url          = {https://doi.org/10.5281/zenodo.20058613},
}

@inproceedings{allenWaterIce,
       author = {{Allen}, Alicia and {Trilling}, David and {Stansberry}, John and {Hilbert}, Bryan and {Strauss}, Ryder and {Thomas}, Cristina and {Holler}, Bryan and {Mueller}, Michael and {Rivkin}, Andrew},
        title = "{Identifying Water Ice in the Asteroid Belt with JWST}",
    booktitle = {EPSC-DPS Joint Meeting 2025},
         year = 2025,
       volume = {2025},
        month = sep,
          eid = {EPSC-DPS2025-1062},
        pages = {EPSC-DPS2025-1062},
          doi = {10.5194/epsc-dps2025-1062},
       adsurl = {https://ui.adsabs.harvard.edu/abs/2025epsc.conf.1062A}
}

@ARTICLE{2022ApJ...935..167A,
       author = {{Astropy Collaboration} and {Price-Whelan}, Adrian M. and {Lim}, Pey Lian and {Earl}, Nicholas and {Starkman}, Nathaniel and {Bradley}, Larry and {Shupe}, David L. and {Patil}, Aarya A. and {Corrales}, Lia and {Brasseur}, C.~E. and {N{\"o}the}, Maximilian and {Donath}, Axel and {Tollerud}, Erik and {Morris}, Brett M. and {Ginsburg}, Adam and {Vaher}, Eero and {Weaver}, Benjamin A. and {Tocknell}, James and {Jamieson}, William and {van Kerkwijk}, Marten H. and {Robitaille}, Thomas P. and {Merry}, Bruce and {Bachetti}, Matteo and {G{\"u}nther}, H. Moritz and {Aldcroft}, Thomas L. and {Alvarado-Montes}, Jaime A. and {Archibald}, Anne M. and {B{\'o}di}, Attila and {Bapat}, Shreyas and {Barentsen}, Geert and {Baz{\'a}n}, Juanjo and {Biswas}, Manish and {Boquien}, M{\'e}d{\'e}ric and {Burke}, D.~J. and {Cara}, Daria and {Cara}, Mihai and {Conroy}, Kyle E. and {Conseil}, Simon and {Craig}, Matthew W. and {Cross}, Robert M. and {Cruz}, Kelle L. and {D'Eugenio}, Francesco and {Dencheva}, Nadia and {Devillepoix}, Hadrien A.~R. and {Dietrich}, J{\"o}rg P. and {Eigenbrot}, Arthur Davis and {Erben}, Thomas and {Ferreira}, Leonardo and {Foreman-Mackey}, Daniel and {Fox}, Ryan and {Freij}, Nabil and {Garg}, Suyog and {Geda}, Robel and {Glattly}, Lauren and {Gondhalekar}, Yash and {Gordon}, Karl D. and {Grant}, David and {Greenfield}, Perry and {Groener}, Austen M. and {Guest}, Steve and {Gurovich}, Sebastian and {Handberg}, Rasmus and {Hart}, Akeem and {Hatfield-Dodds}, Zac and {Homeier}, Derek and {Hosseinzadeh}, Griffin and {Jenness}, Tim and {Jones}, Craig K. and {Joseph}, Prajwel and {Kalmbach}, J. Bryce and {Karamehmetoglu}, Emir and {Ka{\l}uszy{\'n}ski}, Miko{\l}aj and {Kelley}, Michael S.~P. and {Kern}, Nicholas and {Kerzendorf}, Wolfgang E. and {Koch}, Eric W. and {Kulumani}, Shankar and {Lee}, Antony and {Ly}, Chun and {Ma}, Zhiyuan and {MacBride}, Conor and {Maljaars}, Jakob M. and {Muna}, Demitri and {Murphy}, N.~A. and {Norman}, Henrik and {O'Steen}, Richard and {Oman}, Kyle A. and {Pacifici}, Camilla and {Pascual}, Sergio and {Pascual-Granado}, J. and {Patil}, Rohit R. and {Perren}, Gabriel I. and {Pickering}, Timothy E. and {Rastogi}, Tanuj and {Roulston}, Benjamin R. and {Ryan}, Daniel F. and {Rykoff}, Eli S. and {Sabater}, Jose and {Sakurikar}, Parikshit and {Salgado}, Jes{\'u}s and {Sanghi}, Aniket and {Saunders}, Nicholas and {Savchenko}, Volodymyr and {Schwardt}, Ludwig and {Seifert-Eckert}, Michael and {Shih}, Albert Y. and {Jain}, Anany Shrey and {Shukla}, Gyanendra and {Sick}, Jonathan and {Simpson}, Chris and {Singanamalla}, Sudheesh and {Singer}, Leo P. and {Singhal}, Jaladh and {Sinha}, Manodeep and {Sip{\H{o}}cz}, Brigitta M. and {Spitler}, Lee R. and {Stansby}, David and {Streicher}, Ole and {{\v{S}}umak}, Jani and {Swinbank}, John D. and {Taranu}, Dan S. and {Tewary}, Nikita and {Tremblay}, Grant R. and {de Val-Borro}, Miguel and {Van Kooten}, Samuel J. and {Vasovi{\'c}}, Zlatan and {Verma}, Shresth and {de Miranda Cardoso}, Jos{\'e} Vin{\'\i}cius and {Williams}, Peter K.~G. and {Wilson}, Tom J. and {Winkel}, Benjamin and {Wood-Vasey}, W.~M. and {Xue}, Rui and {Yoachim}, Peter and {Zhang}, Chen and {Zonca}, Andrea and {Astropy Project Contributors}},
        title = "{The Astropy Project: Sustaining and Growing a Community-oriented Open-source Project and the Latest Major Release (v5.0) of the Core Package}",
      journal = {\apj},
         year = 2022,
        month = aug,
       volume = {935},
       number = {2},
          eid = {167},
        pages = {167},
          doi = {10.3847/1538-4357/ac7c74},
archivePrefix = {arXiv},
       eprint = {2206.14220},
 primaryClass = {astro-ph.IM},
       adsurl = {https://ui.adsabs.harvard.edu/abs/2022ApJ...935..167A}
}

@ARTICLE{2018AJ....156..123A,
       author = {{Astropy Collaboration} and {Price-Whelan}, A.~M. and {Sip{\H{o}}cz}, B.~M. and {G{\"u}nther}, H.~M. and {Lim}, P.~L. and {Crawford}, S.~M. and {Conseil}, S. and {Shupe}, D.~L. and {Craig}, M.~W. and {Dencheva}, N. and {Ginsburg}, A. and {VanderPlas}, J.~T. and {Bradley}, L.~D. and {P{\'e}rez-Su{\'a}rez}, D. and {de Val-Borro}, M. and {Aldcroft}, T.~L. and {Cruz}, K.~L. and {Robitaille}, T.~P. and {Tollerud}, E.~J. and {Ardelean}, C. and {Babej}, T. and {Bach}, Y.~P. and {Bachetti}, M. and {Bakanov}, A.~V. and {Bamford}, S.~P. and {Barentsen}, G. and {Barmby}, P. and {Baumbach}, A. and {Berry}, K.~L. and {Biscani}, F. and {Boquien}, M. and {Bostroem}, K.~A. and {Bouma}, L.~G. and {Brammer}, G.~B. and {Bray}, E.~M. and {Breytenbach}, H. and {Buddelmeijer}, H. and {Burke}, D.~J. and {Calderone}, G. and {Cano Rodr{\'\i}guez}, J.~L. and {Cara}, M. and {Cardoso}, J.~V.~M. and {Cheedella}, S. and {Copin}, Y. and {Corrales}, L. and {Crichton}, D. and {D'Avella}, D. and {Deil}, C. and {Depagne}, {\'E}. and {Dietrich}, J.~P. and {Donath}, A. and {Droettboom}, M. and {Earl}, N. and {Erben}, T. and {Fabbro}, S. and {Ferreira}, L.~A. and {Finethy}, T. and {Fox}, R.~T. and {Garrison}, L.~H. and {Gibbons}, S.~L.~J. and {Goldstein}, D.~A. and {Gommers}, R. and {Greco}, J.~P. and {Greenfield}, P. and {Groener}, A.~M. and {Grollier}, F. and {Hagen}, A. and {Hirst}, P. and {Homeier}, D. and {Horton}, A.~J. and {Hosseinzadeh}, G. and {Hu}, L. and {Hunkeler}, J.~S. and {Ivezi{\'c}}, {\v{Z}}. and {Jain}, A. and {Jenness}, T. and {Kanarek}, G. and {Kendrew}, S. and {Kern}, N.~S. and {Kerzendorf}, W.~E. and {Khvalko}, A. and {King}, J. and {Kirkby}, D. and {Kulkarni}, A.~M. and {Kumar}, A. and {Lee}, A. and {Lenz}, D. and {Littlefair}, S.~P. and {Ma}, Z. and {Macleod}, D.~M. and {Mastropietro}, M. and {McCully}, C. and {Montagnac}, S. and {Morris}, B.~M. and {Mueller}, M. and {Mumford}, S.~J. and {Muna}, D. and {Murphy}, N.~A. and {Nelson}, S. and {Nguyen}, G.~H. and {Ninan}, J.~P. and {N{\"o}the}, M. and {Ogaz}, S. and {Oh}, S. and {Parejko}, J.~K. and {Parley}, N. and {Pascual}, S. and {Patil}, R. and {Patil}, A.~A. and {Plunkett}, A.~L. and {Prochaska}, J.~X. and {Rastogi}, T. and {Reddy Janga}, V. and {Sabater}, J. and {Sakurikar}, P. and {Seifert}, M. and {Sherbert}, L.~E. and {Sherwood-Taylor}, H. and {Shih}, A.~Y. and {Sick}, J. and {Silbiger}, M.~T. and {Singanamalla}, S. and {Singer}, L.~P. and {Sladen}, P.~H. and {Sooley}, K.~A. and {Sornarajah}, S. and {Streicher}, O. and {Teuben}, P. and {Thomas}, S.~W. and {Tremblay}, G.~R. and {Turner}, J.~E.~H. and {Terr{\'o}n}, V. and {van Kerkwijk}, M.~H. and {de la Vega}, A. and {Watkins}, L.~L. and {Weaver}, B.~A. and {Whitmore}, J.~B. and {Woillez}, J. and {Zabalza}, V. and {Astropy Contributors}},
        title = "{The Astropy Project: Building an Open-science Project and Status of the v2.0 Core Package}",
      journal = {\aj},
         year = 2018,
        month = sep,
       volume = {156},
       number = {3},
          eid = {123},
        pages = {123},
          doi = {10.3847/1538-3881/aabc4f},
archivePrefix = {arXiv},
       eprint = {1801.02634},
 primaryClass = {astro-ph.IM},
       adsurl = {https://ui.adsabs.harvard.edu/abs/2018AJ....156..123A}
}

@ARTICLE{2013A&A...558A..33A,
       author = {{Astropy Collaboration} and {Robitaille}, Thomas P. and
         {Tollerud}, Erik J. and {Greenfield}, Perry and {Droettboom}, Michael and
         {Bray}, Erik and {Aldcroft}, Tom and {Davis}, Matt and
         {Ginsburg}, Adam and {Price-Whelan}, Adrian M. and
         {Kerzendorf}, Wolfgang E. and {Conley}, Alexander and {Crighton}, Neil and
         {Barbary}, Kyle and {Muna}, Demitri and {Ferguson}, Henry and
         {Grollier}, Fr{\'e}d{\'e}ric and {Parikh}, Madhura M. and
         {Nair}, Prasanth H. and {Unther}, Hans M. and {Deil}, Christoph and
         {Woillez}, Julien and {Conseil}, Simon and {Kramer}, Roban and
         {Turner}, James E.~H. and {Singer}, Leo and {Fox}, Ryan and
         {Weaver}, Benjamin A. and {Zabalza}, Victor and {Edwards}, Zachary I. and
         {Azalee Bostroem}, K. and {Burke}, D.~J. and {Casey}, Andrew R. and
         {Crawford}, Steven M. and {Dencheva}, Nadia and {Ely}, Justin and
         {Jenness}, Tim and {Labrie}, Kathleen and {Lim}, Pey Lian and
         {Pierfederici}, Francesco and {Pontzen}, Andrew and {Ptak}, Andy and
         {Refsdal}, Brian and {Servillat}, Mathieu and {Streicher}, Ole},
        title = "{Astropy: A community Python package for astronomy}",
      journal = {\aap},
         year = "2013",
        month = "Oct",
       volume = {558},
          eid = {A33},
        pages = {A33},
          doi = {10.1051/0004-6361/201322068},
archivePrefix = {arXiv},
       eprint = {1307.6212},
 primaryClass = {astro-ph.IM},
       adsurl = {https://ui.adsabs.harvard.edu/abs/2013A&A...558A..33A}
}

@ARTICLE{1996A&AS..117..393B,
       author = {{Bertin}, E. and {Arnouts}, S.},
        title = "{SExtractor: Software for source extraction.}",
      journal = {\aaps},
         year = "1996",
        month = "Jun",
       volume = {117},
        pages = {393-404},
          doi = {10.1051/aas:1996164},
       adsurl = {https://ui.adsabs.harvard.edu/abs/1996A&AS..117..393B}
}

@misc{PIDSURVEYSLICE,
	title = {{JWST} {Program} 5594: JWST Cluster SLICE - Strong LensIng and Cluster Evolution},
	url = {https://www.stsci.edu/jwst-program-info/program/?program=5594},
	publisher = {JWST General Observer Program},
	author = {Mahler, G.},
	year = {2024},
}

@misc{trillingPureParallelSurvey,
	title = {{JWST} {Program} 2211: {A} pure parallel survey of water in the asteroid belt},
	url = {https://www.stsci.edu/jwst-program-info/program/?program=2211},
	publisher = {JWST General Observer Program},
	author = {Trilling, D. E.},
	year = {2023},
}

@misc{trilling_ar_2023,
	title = {{AR} 3701: {Searching} for ultra-faint trans-{Neptunian} objects in archival {NIRCam} calibration data},
	url = {https://www.stsci.edu/jwst/science-execution/program-information?id=3701},
	language = {en},
	urldate = {2026-05-25},
	journal = {STScI},
	author = {Trilling, D. E.},
	year = {2023},
}

@misc{burdanov_ar_2025,
	title = {{AR} 8214: {Mining} {JWST} data for hidden asteroid gems},
	url = {https://www.stsci.edu/jwst/science-execution/program-information?id=8214},
	language = {en},
	urldate = {2026-05-25},
	journal = {STScI},
	author = {Burdanov, A.},
	year = {2025},
}

@ARTICLE{sarc-manual,
       author = {{Micheli}, Marco and {Holler}, Bryan J.},
        title = "{Best practices for obtaining astrometric observations from JWST (274) NIRCam data}",
      journal = {arXiv e-prints},
         year = 2026,
        month = aug,
          eid = {arXiv:2608.26217},
        pages = {arXiv:2608.26217},
          doi = {10.48550/arXiv.2608.26217},
archivePrefix = {arXiv},
       eprint = {2608.26217},
 primaryClass = {astro-ph.IM},
       adsurl = {https://ui.adsabs.harvard.edu/abs/2026arXiv260826217M}
}

@software{webbPSF,
       author = {{Perrin}, Marshall D. and {Long}, Joseph and {Sivaramakrishnan}, Anand and {Lajoie}, Charles-Phillipe and {Elliot}, Erin and {Pueyo}, Laurent and {Albert}, Loic},
        title = "{WebbPSF: James Webb Space Telescope PSF Simulation Tool}",
 howpublished = {Astrophysics Source Code Library, record ascl:1504.007},
         year = 2015,
        month = apr,
          eid = {ascl:1504.007},
archivePrefix = {ascl},
       eprint = {1504.007},
       adsurl = {https://ui.adsabs.harvard.edu/abs/2015ascl.soft04007P}
}

@ARTICLE{gaiaDR3,
       author = {{Gaia Collaboration} and {Vallenari}, A. and {Brown}, A.~G.~A. and {Prusti}, T. and {de Bruijne}, J.~H.~J. and {Arenou}, F. and {Babusiaux}, C. and {Biermann}, M. and {Creevey}, O.~L. and {Ducourant}, C. and {Evans}, D.~W. and {Eyer}, L. and {Guerra}, R. and {Hutton}, A. and {Jordi}, C. and {Klioner}, S.~A. and {Lammers}, U.~L. and {Lindegren}, L. and {Luri}, X. and {Mignard}, F. and {Panem}, C. and {Pourbaix}, D. and {Randich}, S. and {Sartoretti}, P. and {Soubiran}, C. and {Tanga}, P. and {Walton}, N.~A. and {Bailer-Jones}, C.~A.~L. and {Bastian}, U. and {Drimmel}, R. and {Jansen}, F. and {Katz}, D. and {Lattanzi}, M.~G. and {van Leeuwen}, F. and {Bakker}, J. and {Cacciari}, C. and {Casta{\~n}eda}, J. and {De Angeli}, F. and {Fabricius}, C. and {Fouesneau}, M. and {Fr{\'e}mat}, Y. and {Galluccio}, L. and {Guerrier}, A. and {Heiter}, U. and {Masana}, E. and {Messineo}, R. and {Mowlavi}, N. and {Nicolas}, C. and {Nienartowicz}, K. and {Pailler}, F. and {Panuzzo}, P. and {Riclet}, F. and {Roux}, W. and {Seabroke}, G.~M. and {Sordo{\o}rcit}, R. and {Th{\'e}venin}, F. and {Gracia-Abril}, G. and {Portell}, J. and {Teyssier}, D. and {Altmann}, M. and {Andrae}, R. and {Audard}, M. and {Bellas-Velidis}, I. and {Benson}, K. and {Berthier}, J. and {Blomme}, R. and {Burgess}, P.~W. and {Busonero}, D. and {Busso}, G. and {C{\'a}novas}, H. and {Carry}, B. and {Cellino}, A. and {Cheek}, N. and {Clementini}, G. and {Damerdji}, Y. and {Davidson}, M. and {de Teodoro}, P. and {Nu{\~n}ez Campos}, M. and {Delchambre}, L. and {Dell'Oro}, A. and {Esquej}, P. and {Fern{\'a}ndez-Hern{\'a}ndez}, J. and {Fraile}, E. and {Garabato}, D. and {Garc{\'\i}a-Lario}, P. and {Gosset}, E. and {Haigron}, R. and {Halbwachs}, J. -L. and {Hambly}, N.~C. and {Harrison}, D.~L. and {Hern{\'a}ndez}, J. and {Hestroffer}, D. and {Hodgkin}, S.~T. and {Holl}, B. and {Jan{\ss}en}, K. and {Jevardat de Fombelle}, G. and {Jordan}, S. and {Krone-Martins}, A. and {Lanzafame}, A.~C. and {L{\"o}ffler}, W. and {Marchal}, O. and {Marrese}, P.~M. and {Moitinho}, A. and {Muinonen}, K. and {Osborne}, P. and {Pancino}, E. and {Pauwels}, T. and {Recio-Blanco}, A. and {Reyl{\'e}}, C. and {Riello}, M. and {Rimoldini}, L. and {Roegiers}, T. and {Rybizki}, J. and {Sarro}, L.~M. and {Siopis}, C. and {Smith}, M. and {Sozzetti}, A. and {Utrilla}, E. and {van Leeuwen}, M. and {Abbas}, U. and {{\'A}brah{\'a}m}, P. and {Abreu Aramburu}, A. and {Aerts}, C. and {Aguado}, J.~J. and {Ajaj}, M. and {Aldea-Montero}, F. and {Altavilla}, G. and {{\'A}lvarez}, M.~A. and {Alves}, J. and {Anders}, F. and {Anderson}, R.~I. and {Anglada Varela}, E. and {Antoja}, T. and {Baines}, D. and {Baker}, S.~G. and {Balaguer-N{\'u}{\~n}ez}, L. and {Balbinot}, E. and {Balog}, Z. and {Barache}, C. and {Barbato}, D. and {Barros}, M. and {Barstow}, M.~A. and {Bartolom{\'e}}, S. and {Bassilana}, J. -L. and {Bauchet}, N. and {Becciani}, U. and {Bellazzini}, M. and {Berihuete}, A. and {Bernet}, M. and {Bertone}, S. and {Bianchi}, L. and {Binnenfeld}, A. and {Blanco-Cuaresma}, S. and {Blazere}, A. and {Boch}, T. and {Bombrun}, A. and {Bossini}, D. and {Bouquillon}, S. and {Bragaglia}, A. and {Bramante}, L. and {Breedt}, E. and {Bressan}, A. and {Brouillet}, N. and {Brugaletta}, E. and {Bucciarelli}, B. and {Burlacu}, A. and {Butkevich}, A.~G. and {Buzzi}, R. and {Caffau}, E. and {Cancelliere}, R. and {Cantat-Gaudin}, T. and {Carballo}, R. and {Carlucci}, T. and {Carnerero}, M.~I. and {Carrasco}, J.~M. and {Casamiquela}, L. and {Castellani}, M. and {Castro-Ginard}, A. and {Chaoul}, L. and {Charlot}, P. and {Chemin}, L. and {Chiaramida}, V. and {Chiavassa}, A. and {Chornay}, N. and {Comoretto}, G. and {Contursi}, G. and {Cooper}, W.~J. and {Cornez}, T. and {Cowell}, S. and {Crifo}, F. and {Cropper}, M. and {Crosta}, M. and {Crowley}, C. and {Dafonte}, C. and {Dapergolas}, A. and {David}, M. and {David}, P. and {de Laverny}, P. and {De Luise}, F. and {De March}, R. and {De Ridder}, J. and {de Souza}, R. and {de Torres}, A. and {del Peloso}, E.~F. and {del Pozo}, E. and {Delbo}, M. and {Delgado}, A. and {Delisle}, J. -B. and {Demouchy}, C. and {Dharmawardena}, T.~E. and {Di Matteo}, P. and {Diakite}, S. and {Diener}, C. and {Distefano}, E. and {Dolding}, C. and {Edvardsson}, B. and {Enke}, H. and {Fabre}, C. and {Fabrizio}, M. and {Faigler}, S. and {Fedorets}, G. and {Fernique}, P. and {Fienga}, A. and {Figueras}, F. and {Fournier}, Y. and {Fouron}, C. and {Fragkoudi}, F. and {Gai}, M. and {Garcia-Gutierrez}, A. and {Garcia-Reinaldos}, M. and {Garc{\'\i}a-Torres}, M. and {Garofalo}, A. and {Gavel}, A. and {Gavras}, P. and {Gerlach}, E. and {Geyer}, R. and {Giacobbe}, P. and {Gilmore}, G. and {Girona}, S. and {Giuffrida}, G. and {Gomel}, R. and {Gomez}, A. and {Gonz{\'a}lez-N{\'u}{\~n}ez}, J. and {Gonz{\'a}lez-Santamar{\'\i}a}, I. and {Gonz{\'a}lez-Vidal}, J.~J. and {Granvik}, M. and {Guillout}, P. and {Guiraud}, J. and {Guti{\'e}rrez-S{\'a}nchez}, R. and {Guy}, L.~P. and {Hatzidimitriou}, D. and {Hauser}, M. and {Haywood}, M. and {Helmer}, A. and {Helmi}, A. and {Sarmiento}, M.~H. and {Hidalgo}, S.~L. and {Hilger}, T. and {H{\l}adczuk}, N. and {Hobbs}, D. and {Holland}, G. and {Huckle}, H.~E. and {Jardine}, K. and {Jasniewicz}, G. and {Jean-Antoine Piccolo}, A. and {Jim{\'e}nez-Arranz}, {\'O}. and {Jorissen}, A. and {Juaristi Campillo}, J. and {Julbe}, F. and {Karbevska}, L. and {Kervella}, P. and {Khanna}, S. and {Kontizas}, M. and {Kordopatis}, G. and {Korn}, A.~J. and {K{\'o}sp{\'a}l}, {\'A} and {Kostrzewa-Rutkowska}, Z. and {Kruszy{\'n}ska}, K. and {Kun}, M. and {Laizeau}, P. and {Lambert}, S. and {Lanza}, A.~F. and {Lasne}, Y. and {Le Campion}, J. -F. and {Lebreton}, Y. and {Lebzelter}, T. and {Leccia}, S. and {Leclerc}, N. and {Lecoeur-Taibi}, I. and {Liao}, S. and {Licata}, E.~L. and {Lindstr{\o}m}, H.~E.~P. and {Lister}, T.~A. and {Livanou}, E. and {Lobel}, A. and {Lorca}, A. and {Loup}, C. and {Madrero Pardo}, P. and {Magdaleno Romeo}, A. and {Managau}, S. and {Mann}, R.~G. and {Manteiga}, M. and {Marchant}, J.~M. and {Marconi}, M. and {Marcos}, J. and {Marcos Santos}, M.~M.~S. and {Mar{\'\i}n Pina}, D. and {Marinoni}, S. and {Marocco}, F. and {Marshall}, D.~J. and {Polo}, L. Martin and {Mart{\'\i}n-Fleitas}, J.~M. and {Marton}, G. and {Mary}, N. and {Masip}, A. and {Massari}, D. and {Mastrobuono-Battisti}, A. and {Mazeh}, T. and {McMillan}, P.~J. and {Messina}, S. and {Michalik}, D. and {Millar}, N.~R. and {Mints}, A. and {Molina}, D. and {Molinaro}, R. and {Moln{\'a}r}, L. and {Monari}, G. and {Mongui{\'o}}, M. and {Montegriffo}, P. and {Montero}, A. and {Mor}, R. and {Mora}, A. and {Morbidelli}, R. and {Morel}, T. and {Morris}, D. and {Muraveva}, T. and {Murphy}, C.~P. and {Musella}, I. and {Nagy}, Z. and {Noval}, L. and {Oca{\~n}a}, F. and {Ogden}, A. and {Ordenovic}, C. and {Osinde}, J.~O. and {Pagani}, C. and {Pagano}, I. and {Palaversa}, L. and {Palicio}, P.~A. and {Pallas-Quintela}, L. and {Panahi}, A. and {Payne-Wardenaar}, S. and {Pe{\~n}alosa Esteller}, X. and {Penttil{\"a}}, A. and {Pichon}, B. and {Piersimoni}, A.~M. and {Pineau}, F. -X. and {Plachy}, E. and {Plum}, G. and {Poggio}, E. and {Pr{\v{s}}a}, A. and {Pulone}, L. and {Racero}, E. and {Ragaini}, S. and {Rainer}, M. and {Raiteri}, C.~M. and {Rambaux}, N. and {Ramos}, P. and {Ramos-Lerate}, M. and {Re Fiorentin}, P. and {Regibo}, S. and {Richards}, P.~J. and {Rios Diaz}, C. and {Ripepi}, V. and {Riva}, A. and {Rix}, H. -W. and {Rixon}, G. and {Robichon}, N. and {Robin}, A.~C. and {Robin}, C. and {Roelens}, M. and {Rogues}, H.~R.~O. and {Rohrbasser}, L. and {Romero-G{\'o}mez}, M. and {Rowell}, N. and {Royer}, F. and {Ruz Mieres}, D. and {Rybicki}, K.~A. and {Sadowski}, G. and {S{\'a}ez N{\'u}{\~n}ez}, A. and {Sagrist{\`a} Sell{\'e}s}, A. and {Sahlmann}, J. and {Salguero}, E. and {Samaras}, N. and {Sanchez Gimenez}, V. and {Sanna}, N. and {Santove{\~n}a}, R. and {Sarasso}, M. and {Schultheis}, M. and {Sciacca}, E. and {Segol}, M. and {Segovia}, J.~C. and {S{\'e}gransan}, D. and {Semeux}, D. and {Shahaf}, S. and {Siddiqui}, H.~I. and {Siebert}, A. and {Siltala}, L. and {Silvelo}, A. and {Slezak}, E. and {Slezak}, I. and {Smart}, R.~L. and {Snaith}, O.~N. and {Solano}, E. and {Solitro}, F. and {Souami}, D. and {Souchay}, J. and {Spagna}, A. and {Spina}, L. and {Spoto}, F. and {Steele}, I.~A. and {Steidelm{\"u}ller}, H. and {Stephenson}, C.~A. and {S{\"u}veges}, M. and {Surdej}, J. and {Szabados}, L. and {Szegedi-Elek}, E. and {Taris}, F. and {Taylo}, M.~B. and {Teixeira}, R. and {Tolomei}, L. and {Tonello}, N. and {Torra}, F. and {Torra}, J. and {Torralba Elipe}, G. and {Trabucchi}, M. and {Tsounis}, A.~T. and {Turon}, C. and {Ulla}, A. and {Unger}, N. and {Vaillant}, M.~V. and {van Dillen}, E. and {van Reeven}, W. and {Vanel}, O. and {Vecchiato}, A. and {Viala}, Y. and {Vicente}, D. and {Voutsinas}, S. and {Weiler}, M. and {Wevers}, T. and {Wyrzykowski}, L. and {Yoldas}, A. and {Yvard}, P. and {Zhao}, H. and {Zorec}, J. and {Zucker}, S. and {Zwitter}, T.},
        title = "{Gaia Data Release 3: Summary of the content and survey properties}",
      journal = {arXiv e-prints},
         year = 2022,
        month = jul,
          eid = {arXiv:2208.00211},
        pages = {arXiv:2208.00211},
archivePrefix = {arXiv},
       eprint = {2208.00211},
 primaryClass = {astro-ph.GA},
       adsurl = {https://ui.adsabs.harvard.edu/abs/2022arXiv220800211G}
}

@software{rampSlicerv1.1.1,
  author       = {Girmenia, Anthony},
  title        = {JWST-Ramp-Slicer},
  month        = sep,
  year         = 2026,
  publisher    = {Zenodo},
  version      = {v1.1.1},
  doi          = {10.5281/zenodo.22664430},
  url          = {https://doi.org/10.5281/zenodo.22664430},
}
\bibliographystyle{aasjournalv7}


\end{document}